\documentclass[twocolumn]{aastex631}

\usepackage{changepage}
\usepackage{rotating}

\begin{document}
\title{In Search of Decay: An Analysis of Transit Times of Hot Jupiters in Main Sequence and Post-Main Sequence Systems} 

\author[0009-0005-8360-9173]{Noah Sodickson}
\affiliation{Original Science Research Program, Mamaroneck High School, 1000 W. Boston Post Road, Mamaroneck, NY 10543, USA}

\author[0000-0003-4976-9980]{Samuel Grunblatt}
\affiliation{Department of Physics and Astronomy, Johns Hopkins University, 3400 N. Charles Street, Baltimore, MD 21218, USA}
\affiliation{Department of Physics and Astronomy, The University of Alabama, 514 University Blvd., Tuscaloosa, AL 35487, USA}

\begin{abstract}

Tidal interactions are one of the primary drivers of orbital evolution for massive planets with short orbital periods. Tidal dissipation within host stars can cause the orbits of such planets to decay. However, the mechanisms of tidal dissipation are difficult to probe. Generally, tidal dissipation is parameterized by the modified stellar tidal quality factor, or $Q_{*}^{'}$, but the lack of observational evidence of orbital decay to confirm dissipation theories has resulted in orders of magnitude of uncertainty in $Q_{*}^{'}$. We present a new transit timing analysis of 54 systems with varying stellar evolutionary states in an attempt to search for orbital decay across multiple stages of stellar evolution. For each system, we obtained mid-transit times from new TESS data and evaluated potential departures from a linear ephemeris using the Bayesian Information Criterion. We then determined tidal quality factors using widely tested theoretical relations. Of the systems studied, 25 showed evidence of a decrease in orbital period over time and 8 showed evidence of a decrease that was inconsistent with 0 by three standard deviations: CoRoT-2, TrES-5, WASP-4, WASP-12, WASP-19, WASP-45, WASP-99, and XO-3. However, the significance of some of these detections may be influenced by unreliable transit time measurements. Similarly, we see that the lower limit on $Q_{*}^{'}$ for evolved systems is marginally lower than it is for the sample of all systems, though both limits are orders of magnitude below the expected theoretical values for both samples. These new constraints on transit times and $Q_{*}^{'}$ values will help to narrow the search for orbital decay in the future and place important constraints on current theories of star-planet interactions.

\end{abstract}


\section{Introduction} \label{sec:intro}
The first hot Jupiter was discovered by \citet{Mayor1995}. Since then, hot Jupiters, or planets with a mass around that of Jupiter and a period of less than 10 days, have become some of the most commonly studied exoplanets. It is well known that tidal interactions between hot Jupiters and their host stars play a major role in planetary evolution. These interactions are driven by tidal dissipation, or the process by which tidal forces create friction within stars. Tidal dissipation can lead to a decrease in the eccentricities of orbits, or spin-orbit synchronization \citep{Hut1980}. When planets orbit faster than their host stars rotate, which is true of almost all hot Jupiter systems, it can also lead to tidally-driven orbital decay.
 
Unfortunately, the mechanisms by which tidal dissipation occurs in hot Jupiter systems and the timescales of orbital decay are not well understood. Tidal dissipation was originally studied in binary star systems, and the breaking of equilibrium tides was the first proposed mechanism. However, it has since been shown that dynamical tides, or the excitation of oscillations within a star, can be more dominant, especially during later stages of stellar evolution \citep{Zahn1975, Zahn1977}. The same has been found to be the case in hot Jupiter systems \citep{Lubow1997}. Dynamical tides can be further separated into two categories. The first are inertial wave-driven tides which are dissipated in the convective zones of stars \citep{Ogilvie2004}. The second are internal gravity wave-driven tides which are dissipated in the radiative zones of stars and thought to be the dominant mechanism of orbital decay in hot Jupiter systems. Internal gravity waves can be dissipated by multiple phenomena, but they are most efficiently dissipated by wave breaking, which occurs when the planet exceeds a critical mass. This critical mass strongly depends on stellar mass and age, rapidly decreasing to much below the mass of Jupiter as the star transitions off the main sequence and then increasing back up to approximately the mass of Jupiter as shown by Figure 9 in \citet{Barker2020}. As a result the effectiveness of tidal dissipation can change drastically throughout a star’s evolution \citep{Barker2010}.

Hot Jupiters are especially likely to experience tidal interactions with their host stars because of their large masses and small orbital separations. It was once thought that orbital decay resulted in the destabilization of almost all close-in hot Jupiters and other close-in exoplanets before their stars leave the main sequence \citep{Schlaufman2013}, though counterexamples have now been discovered, \citep[e.g.][]{Chontos2019, Schanche2020, Grunblatt2022}. Strong evidence of orbital decay has been observed for WASP-12 b, which is thought to be orbiting either a main-sequence star or a slightly evolved star, and Kepler-1658 b, which is orbiting a clearly evolved star. Kepler-1658 b was found to have a faster decay rate, which is consistent with the fact that evolved stars are thought to be more dissipative \citep{Patra2017, Bailey2019, Turner2021, Vissapragada2022}.

The modified stellar tidal quality factor, denoted in this work as $Q_{*}^{'}$, is generally used to parameterize tidal dissipation, where a lower value indicates more efficient dissipation. $Q_{*}^{'}$ is proportional to the ratio of energy stored in tides during one orbit to the energy lost to tidal dissipation in one orbit. $Q_{*}^{'}$ is given by the following equations:
\begin{equation} \label{eq:eqone}
Q^{-1} = \frac{1}{2 \pi E_{0}} \oint(-\frac{dE}{dt}) dt
\end{equation}
\begin{equation} \label{eq:eqtwo}
Q_{*}^{'} = \frac{3Q}{2k_{2}}
\end{equation}
Where $2 \pi E_{0}$ is the amount of tidal energy stored in one orbit, $-\frac{dE}{dt}$ is the rate of energy loss due to tidal dissipation, and $k_{2}$ is a dimensionless quantity called Love number that describes the density of the star \citep{Goldreich1966}. Measuring or providing observational constraints on $Q_{*}^{'}$ for systems at different evolutionary states is extremely important because different tidal mechanisms predict $Q_{*}^{'}$ values that differ by orders of magnitude \citep{Barker2020}. 

Observable orbital decay has been suggested for many other hot Jupiters, including XO-3 b and WASP-19 b. One of the most useful recent databases to investigate this was created by \citet{Ivshina2022} (hereafter referred to as IW22). IW22 measured individual transit times and compiled transit  times from the literature for 348 systems, 240 of which were hot Jupiters. One major improvement of the IW22 sample is the inclusion of TESS measurements, which are extending the baseline of transit measurements for the brightest hot Jupiter systems across the sky. Launched in 2018, TESS is the newest telescope that lets us search for orbital decay in many hot Jupiter systems, making a higher level of precision in transit timing measurements possible \citep{Huber2022}. By including TESS measurements at the shortest cadences available, IW22 were able to identify some evidence of orbital decay in 9 systems for which orbital decay had not yet been observed. Additional studies have analyzed transit times using TESS data, some of which incorporated the IW22 times, and further extended the number of orbital decay candidates \citep[e.g.][]{Patra2020, Wang2023, Adams2024}.

In order to address the need for more observations of tidally-driven orbital decay to better understand tidal dissipation, this study aims to perform a comprehensive search for orbital decay in a variety of systems in different stages of stellar evolution by combining transit times measured from the newest 2 minute and 20 second cadence TESS data obtained after the study IW22 with those reported by IW22. For each system, we fit transit times to both a linear and quadratic ephemeris and used Bayesian Information Criterion to determine which model fits better. Then, for systems where a quadratic ephemeris fits well, $Q_{*}^{'}$ and its $1 \sigma$ uncertainties are calculated and reported, and for the remaining systems a lower limit on $Q_{*}^{'}$ is calculated. Finally, whether or not our $Q_{*}^{'}$ values are consistent with both theoretical predictions and previous studies, and the implications this has on stellar structures and dissipation mechanisms are discussed. For some systems, alternative phenomena that could cause an apparent change in orbital period similar to the change caused by tidally-driven orbital decay are also considered.

\section{Methods} \label{sec:methods}

\subsection{Target and Data Selection} \label{subsec:data}
We first compiled a list of evolved systems containing hot Jupiters observed by TESS based on our in-house analysis which determined the host star has likely reached a post-main sequence evolutionary state ($T_{\mathrm{eff}} \lesssim 6000$ K, $R_{*} \gtrsim 2 R_{\odot}$), as planets orbiting evolved stars are thought to be more susceptible to rapid orbital decay. We determined evolutionary states of these stars through analysis of stellar properties provided by the TESS Input Catalog \citep{Stassun2018}. In addition, we added systems from the literature for which orbital decay has been previously studied. Most of these additional systems were main-sequence systems. The selection is not necessarily representative of the known population of hot Jupiters. Our final list consisted of 54 systems, 51 of which contained confirmed hot Jupiters. The remaining 3 have been flagged as TESS objects of interest (TOI).


For each system, we downloaded system parameters from the Exoplanet Follow-up Observing Program database, or ExoFOP. Planet radius, impact parameter, and semi-major axis were used as priors for transit modeling and stellar radius, planet and stellar mass, and semi-major axis were used to calculate $Q_{*}^{'}$. For systems whose stellar and planet masses weren't available on ExoFOP, we found listed stellar and planet masses from the NASA exoplanet archive. Only TOI-2787 didn't have a listed stellar mass, and only TOI-2787, TOI-4436, and K2-161 didn't have listed planet masses. For those systems, we used the average stellar mass of our sample of 1.14 M$_{\odot}$ and the average planet mass of our sample of 846.55 M$_{\Earth}$ when necessary.


Table \ref{tab:tabone} shows the parameters and uncertainties we obtained from ExoFOP and the NASA exoplanet archive.

\begin{table*}
\begin{center}
\caption{System Parameters from ExoFOP \label{tab:tabone}}
\begin{tabular}{lllllllll}
\hline
System & Period & $t_{0}$ & $R_{\star}$ & $R_{p}$ & $M_{\star}$ & $M_{p}$ & $a$ & $b^{b}$ \\
& (days) & (BJD$^{a}$) & (R$_{\odot}$) & (R$_{\Earth}$) & (M$_{\odot}$) & (M$_{\Earth}$) & (AU) & \\
\hline
CoRoT-1 & $1.51 \pm 6.4 \times 10^{-6}$ & $-2840.55$ & $1.29$ & $16.70 \pm 0.9$ & $1.26$ & $327.35 \pm 38.14$ &   & \\
CoRoT-2 & $1.74 \pm 1.0 \times 10^{-6}$ & $-2762.46$ & $0.94 \pm 0.095$ & $16.43 \pm 0.47$ & $0.97 \pm 0.12$ & $1102.82 \pm 69.92$ & $0.03 \pm 0.00076$ & $0.22$ \\
CoRoT-5 & $4.04 \pm 1.9 \times 10^{-6}$ & $-2599.80$ & $1.27$ & $15.56 \pm 0.52$ & $1.21$ & $148.42 \pm 14.94$ & $0.05 \pm 0.00026$ & $0.76$ \\
HAT-P-23 & $1.21 \pm 2.0 \times 10^{-6}$ & $-2147.74$ & $1.15 \pm 0.06$ & $15.33 \pm 1.01$ & $1.08 \pm 0.14$ & $664.24 \pm 35.28$ & $0.02 \pm 0.0002$ & $0.32$ \\
HAT-P-40 & $4.46 \pm 1.0 \times 10^{-5}$ & $2855.87 \pm 0.0011$ & $2.09 \pm 0.098$ & $17.04 \pm 1.91$ & $1.14 \pm 0.15$ & $152.56 \pm 41.32$ &   & \\
HAT-P-60 & $4.79 \pm 2.4 \times 10^{-6}$ & $1360.94$ & $2.18 \pm 0.097$ & $18.28 \pm 0.27$ & $1.25 \pm 0.18$ & $182.43 \pm 12.08$ & $0.06 \pm 0.00017$ & $0.67$ \\
HAT-P-67 & $4.81 \pm 4.3 \times 10^{-7}$ & $-1038.62$ & $2.65 \pm 0.12$ & $23.37 \pm 1.08$ & $1.29 \pm 0.21$ & $108.06 \pm 79.46$ & $0.07 \pm 0.0027$ & $0.12$ \\
HATS-18 & $0.84 \pm 4.7 \times 10^{-7}$ & $1569.54 \pm 0.0002$ & $1.02$ & $14.90 \pm 0.84$ & $1.01$ & $629.14 \pm 24.16$ & $0.02 \pm 0.00098$ & $0.30$ \\
K2-97 & $8.41 \pm 2.3 \times 10^{-5}$ & $722.14$ & $4.71$ & $13.45 \pm 1.23$ & $1.17 \pm 0.19$ & $174.49 \pm 18.75$ & $0.09 \pm 0.0044$ & $0.90$ \\
K2-132 & $9.18 \pm 0.0023$ & $590.15$ & $3.79$ & $14.57 \pm 0.78$ & $1.08 \pm 0.08$ & $155.74 \pm 19.07$ &   & $0.85$ \\
K2-161 & $9.28 \pm 0.0021$ & $587.85$ & $4.23$ & $6.12 \pm 1.63$ & $1.30 \pm 0.15$ &   &   & \\
\hline
\multicolumn{9}{l}{a. Barycentric Julian Date minus 2457000 days.} \\
\multicolumn{9}{l}{b. Transit Impact Parameter.} \\
\multicolumn{9}{l}{(Entire table is available for download in machine readable format.)} \\
\end{tabular}
\end{center}
\end{table*}

We downloaded TESS 2 minute cadence and SPOC light curves for each system, along with 20 second cadence light curves when available, from the Mikulski Archive for Space Telescopes using the \texttt{Astroquery} Python library \citep{Ginsburg2019}. For the few systems on our list observed by the Kepler telescope, we also downloaded Kepler and K2SFF data and analyzed it using an equivalent approach as we used to analyze TESS data. We then flattened and normalized the light curves, and removed outliers by clipping data points beyond five standard deviations from the local median. For light curves with notable variability on time scales larger than the transit period, we removed trends by dividing the light curves by median filtered data using a window size of a quarter of the period. Finally, we used the \texttt{Lightkurve} Python library to combine light curves from multiple sectors to form our final light curves for analysis, so we could fit all of the TESS data for each system with one model \citep{Cardoso2018}.

\subsection{Obtaining Mid Transit Times} \label{subsec:times}
\begin{figure*}[ht!]
\plotone{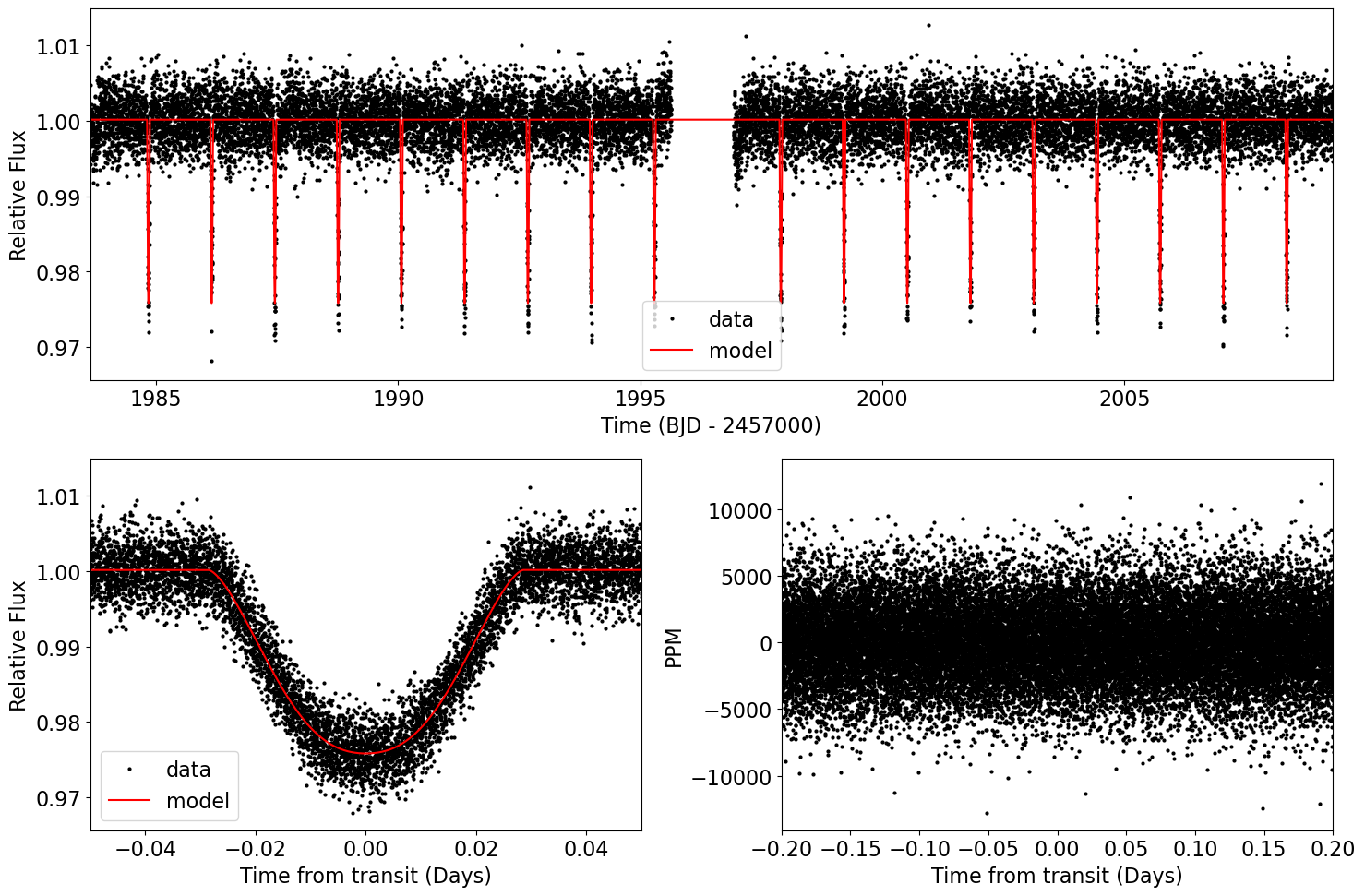}
\caption{\textit{(Top)} The SPOC light curve of TrES-3 from TESS Sector 25, with the best fit transit model over-plotted (red line). The time of each transit was fit independently. \textit{(Bottom Left)} The light curve data and transit model phase-folded at the measured orbital period. \textit{(Bottom Right)} The residuals of the phase-folded light curve after subtracting the transit model, in parts per million (PPM). \label{fig:figone}}
\end{figure*}

\begin{figure*}[ht!]
\plotone{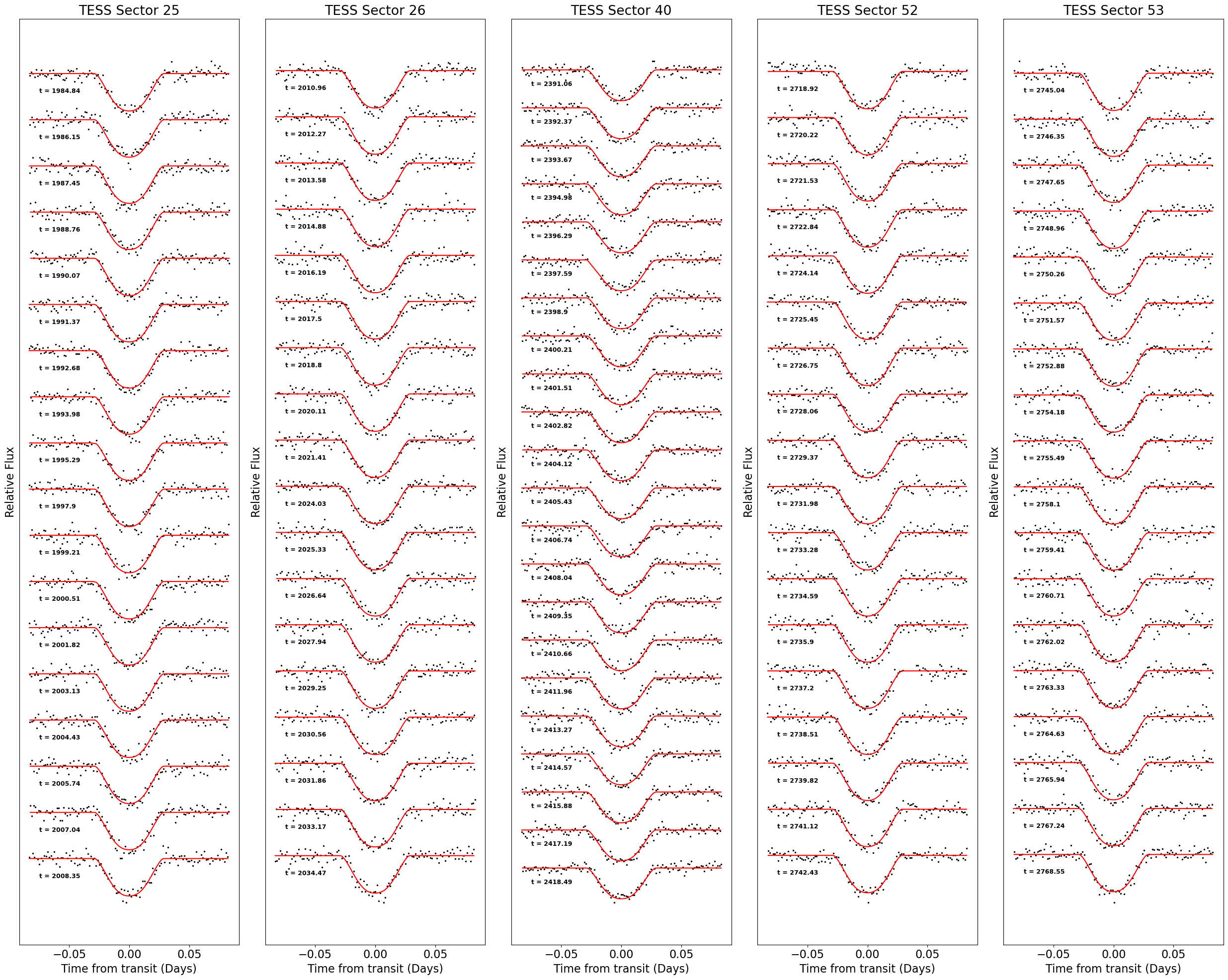}
\caption{Each individually fit transit for TrES-3. The transits are horizontally aligned based on the sector they were observed in, and stacked in order of time. The y-axis is relative flux. \label{fig:figtwo}}
\end{figure*}

To obtain mid transit times for each system, we modeled the light curves using the \texttt{Exoplanet} Python library and the \texttt{PyMC} Python library \citep{Foreman-Mackey2021, Salvatier2015}. The parameters we chose to fit were out of transit mean flux, planet radius, impact parameter, and two quadratic limb darkening coefficients \citep{Kipping2013}. Additionally, we treated each mid-transit time as an independent, normally distributed parameter with a prior determined by the linear ephemeris in ExoFOP. We determined posterior parameter distributions using Markov Chain Monte Carlo (MCMC) modeling, producing two chains of 2000 steps each, discarding the first 1000 steps as tuning steps, to obtain posterior distributions, and recorded the maximum a posteriori transit times. Figure \ref{fig:figone} shows the best fit transit model for the TrES-3 system, which had the most data of any system in our sample, and figure \ref{fig:figtwo} shows the individual transits of the TrES-3 system with the same model overlaid.

To supplement the transit times obtained in this study, we also downloaded the transit times obtained by IW22. IW22 obtained transit times for 382 systems. For those systems in our target list that had data published in IW22, we used their transit times and added the additional transit times we determined from more recently acquired TESS data for our analysis. This selection process is outlined in \S \ref{subsec:data}. Outliers were removed from the set of all transit times by clipping times whose deviations from the ExoFOP ephemeris varied by five standard deviations from the local mean deviation. This removed fewer than 10\% of the transits in our overall database, but ensured accurate fits. 

\subsection{Fitting Linear and Quadratic Ephemerides} \label{subsec:ephemerides}
We fit the transit times we determined along with the previous measurements from IW22 to build a linear ephemeris model and orbital decay timing model for each system. We have listed all of our measured transit times in table \ref{tab:tabfive} in the appendix. The table is available in its entirety in machine readable format. We formulated our linear ephemeris model as follows:
\begin{equation} \label{eq:eqthree}
t_{pred}(E) = t_{0} + PE
\end{equation}
where E is a transit number, $t_{0}$ is the transit epoch, and $P$ is a constant period \citep{Patra2017}. We developed the following equation to determine the epoch of a given observed transit time:
\begin{equation} \label{eq:eqfour}
E = \lfloor (t_{obs} - (t_{obs} \: mod \: P - t_{0} \: mod \: P) - t_{0}) / P + 0.5 \rfloor
\end{equation}
where $t_{obs} \: mod \: P - t_{0} \: mod \: P$ is the deviation from the transit time predicted by a linear ephemeris with $t_{0}$ and $P$. To apply equation \ref{eq:eqfour}, we used the period and epoch present in ExoFOP provided by the NASA Exoplanet Archive before refining them during the modeling process. We formulated our orbital decay model using the following quadratic function:
\begin{equation} \label{eq:eqfive}
t_{pred}(E) = t_{0} + PE + \frac{1}{2} \frac{dP}{dE} E^{2}
\end{equation}
where $\frac{dP}{dE}$ is a period derivative with respect to epoch \citep{Patra2017}. A negative $\frac{dP}{dE}$ would mean a shrinking period, potentially indicative of orbital decay. For the $t_{0}$ and $P$ parameters in both models, we used uniform priors centered around the values from ExoFOP, and for the $\frac{dP}{dE}$ parameter in the orbital decay model, we used a uniform prior centered around 0, assuming a baseline of no period decay. We determined posterior parameter distributions using MCMC modeling, producing four chains with 6000 steps each, discarding the first 3000 steps as tuning steps, and recorded the median model parameters. We also recorded the 16th and 84th model parameter percentiles, which we report as the $1 \sigma$ uncertainties. We flagged any target for which the median $\frac{dP}{dE}$ was inconsistent with 0 by $3 \sigma$ based on the posterior distribution of $\frac{dP}{dE}$ from the MCMC, using the same criterion as IW22.

To compare both models, we used Bayesian Information Criterion, or $BIC$, given by the following equations:
\begin{equation} \label{eq:eqsix}
BIC = \chi^{2} + k \times ln(N)
\end{equation}
\begin{equation} \label{eq:eqseven}
\Delta BIC = BIC_{linear} - BIC_{quadratic}
\end{equation}
A lower $BIC$ indicates a better fit, and $\Delta BIC$ is the $BIC$ for the linear ephemeris model minus $BIC$ for the quadratic ephemeris model. Therefore, a positive $\Delta BIC$ indicates that the quadratic model fits better than the linear model.

We calculated $Q_{*}^{'}$ for every system with a negative $\frac{dP}{dE}$. We used the following equations from \citet{Patra2017} based on the constant lag-angle model from \citet{Goldreich1966}:
\begin{equation} \label{eq:eqeight}
\dot{P} = \frac{1}{P} \frac{dP}{dE}
\end{equation}
\begin{equation} \label{eq:eqnine}
\dot{P} = -\frac{27 \pi}{2Q_{*}^{'}} \frac{M_{p}}{M_{*}} (\frac{R_{*}}{a})^{5}
\end{equation}
where $M_{p}$ and $M_{*}$ are the planet and stellar mass, respectively, $R_{*}$ is the stellar radius, and $a$ is the orbital semi-major axis. We also calculated lower limits on $Q_{*}^{'}$ by also applying equations \ref{eq:eqeight} and \ref{eq:eqnine} to our measured $3 \sigma$ lower errors in $\frac{dP}{dE}$. The majority of systems had either negative $\frac{dP}{dE}$ values or positive $\frac{dP}{dE}$ values within $3 \sigma$ of zero, so we were able to at least constrain $Q_{*}^{'}$ in around $85\%$ systems.

\section{Results} \label{sec:results}

\begin{figure*}[ht!]
\plotone{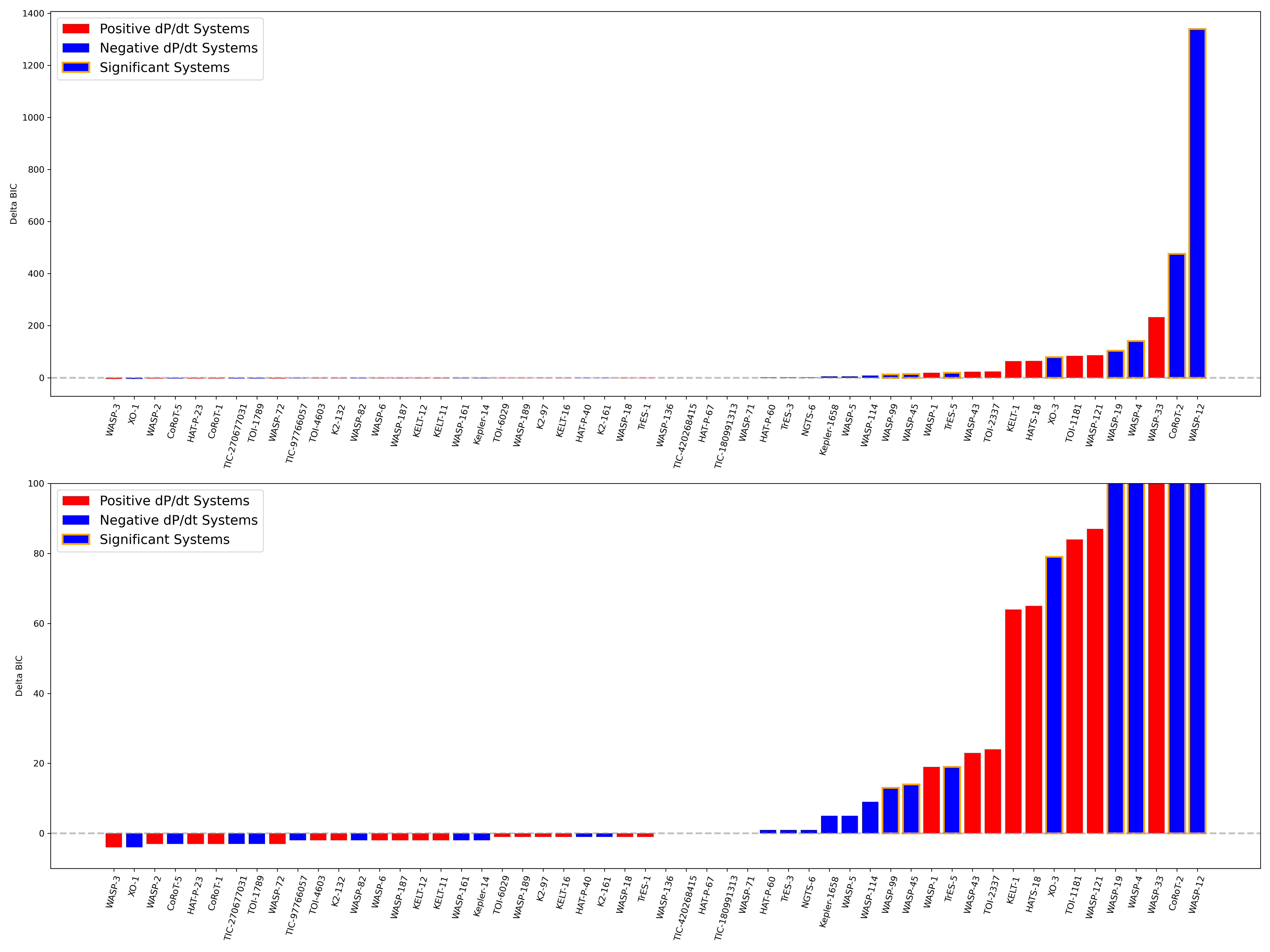}
\caption{All 57 systems sorted by $\Delta BIC$ between a linear transit timing model and a quadratic transit timing model. Systems with a positive $\dot{P}$ are colored blue while systems with a negative $\dot{P}$ are colored red. Furthermore, systems that passed the $3 \sigma$ criterion and had a $\Delta BIC > 10$, and therefore show significant evidence of a decrease in orbital period over time, are outlined in orange. \textit{(Bottom)} Zoomed in to better display the $\Delta BIC$ of most systems \label{fig:figthree}}
\end{figure*}

Figure \ref{fig:figthree} shows the $\Delta BIC$ of every system. Approximately half of all systems have a positive $\Delta BIC$ indicating a better fit to the quadratic model. We chose to flag every system that met the $3 \sigma$ criterion and had a $\Delta BIC > 10$ as having significant evidence of a decrease in orbital period over time. We also discuss systems that met the $3 \sigma$ criterion but had a $\Delta BIC < 10$ below, but consider them to be significantly weaker candidates. Furthermore, we flagged every remaining system whose $\frac{dP}{dE}$ values were inconsistent with 0 by at least $1 \sigma$ and which had a positive $\Delta BIC$ value as having any evidence of a decrease in orbital period over time. However, we note that several of the systems in this population had $Q_{*}^{'} < 10^{4}$ which is lower than the predictions of most stellar models and therefore likely underestimated or not resulting from tidal orbital decay \citep[e.g.][]{Weinberg2024}. Additionally, several of the systems that may show evidence of a period decrease have relatively few effective epochs of data, making it challenging to meaningfully constrain changes in orbital period. Model outputs for our significant systems and systems with any evidence of a nonlinear ephemeris can be found in table \ref{tab:tabtwo}, and $\Delta BIC$ values and $Q_{*}^{'}$ values can be found in table \ref{tab:tabthree}. Additionally, $\Delta BIC$ values and lower limits on $Q_{*}^{'}$ for the remaining systems can be found in table \ref{tab:tabfour}. To ensure results that were as accurate as possible for our most significant systems, we manually obtained system parameters from specific studies and used them to recalculate the $Q_{*}^{'}$ values. We then compared our $Q_{*}^{'}$ values with theoretical values obtained by \citet{Barker2020} (hereafter referred to as B20) and \citet{Weinberg2024} (hereafter referred to as W24), both of which we denote as $Q_{IGW}^{'}$. The two different values represent different theoretical approaches to estimating tidal quality factors. Both studies assumed that dynamical tides, specifically internal gravity waves, are the dominant driver of tidal dissipation. B20 assumes that these waves are fully damped, possibly by wave breaking, which is a strongly nonlinear mechanism. W24, on the other hand, simulates a sea of secondary waves in the stellar core excited by these waves, which is a weakly non-linear mechanism. As a result of B20's assumption that the waves are always fully damped, their values tend to be lower than the values obtained by W24. Additionally, W24 calculated $Q_{IGW}^{'}$ values for stellar models with discrete sets of parameters, so the estimates we use for comparison are for the closest model to the system in question, or the closest two models if ranges for $Q_{IGW}^{'}$ are written. We discuss our most significant systems below.

\begin{figure*}[ht!]
\plotone{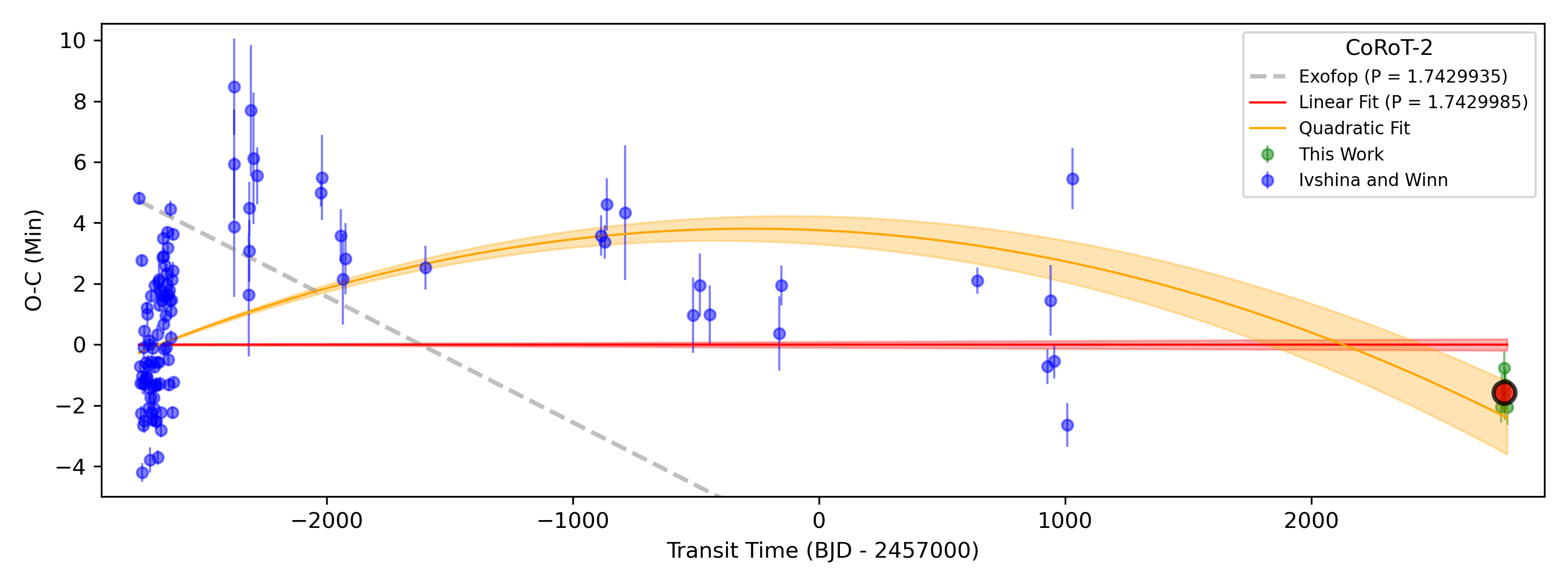}
\caption{Observed minus calculated transit times in minutes as a function of time in days in for CoRoT-2, showing our best fit linear transit timing model and quadratic transit timing model in red and orange, respectively. The blue circles represent transit times obtained from IW22, the green circles represent transit times obtained from this work, and the red circles are the average transit times obtained from this work for each TESS sector and Kepler quarter. The shaded regions around the red and orange lines are $1 \sigma$ uncertainties. Additionally, the dotted line represents the default orbital period reported by the NASA Exoplanet Archive \label{fig:figfour}}
\end{figure*}

\textbf{CoRoT-2:} CoRoT-2 b is a hot Jupiter with a mass of $3.31 \pm 0.16 \: M_{J}$ and a radius of $1.41 \pm 0.06 \: R_{J}$ \citep{Ozturk2019}. We fit 123 timing measurements, 7 of which were newly observed since the IW22 times were published. This gave us a $\dot{P}$ of $-50.32 \pm 2.18 \: ms \: yr^{-1}$, and a $Q_{*}^{'}$ of $(6.33 \pm 0.27) \times 10^{3}$. The quadratic fit was found to be significantly better than the linear fit, with a $\Delta BIC$ of 507. CoRoT-2 was not specifically discussed by B20. However the stellar age estimate of $2.7_{-2.7}^{+3.2} \: Gyr$ obtained by \citet{Bonomo2017} gives a $Q_{IGW}^{'}$ of approximately $2.2 \times 10^{5}$ from figure 8 in B20, which is two orders of magnitude above the value measured in this work. The parameters of the CoRoT-2 system did not closely match any models from table 4 in W24. CoRoT-2 had one of the highest $\Delta BIC$ values of all the systems in this study. However, the IW22 times for CoRoT-2, especially the earlier cluster of times, are known to have under-reported error bars and mistakes due to the conversion between the UTC timing system and the BJD/TDB timing system \citep{Adams2024}. When \citet{Adams2024} corrected these errors, they found insignificant evidence for orbital decay.

\begin{figure*}[ht!]
\plotone{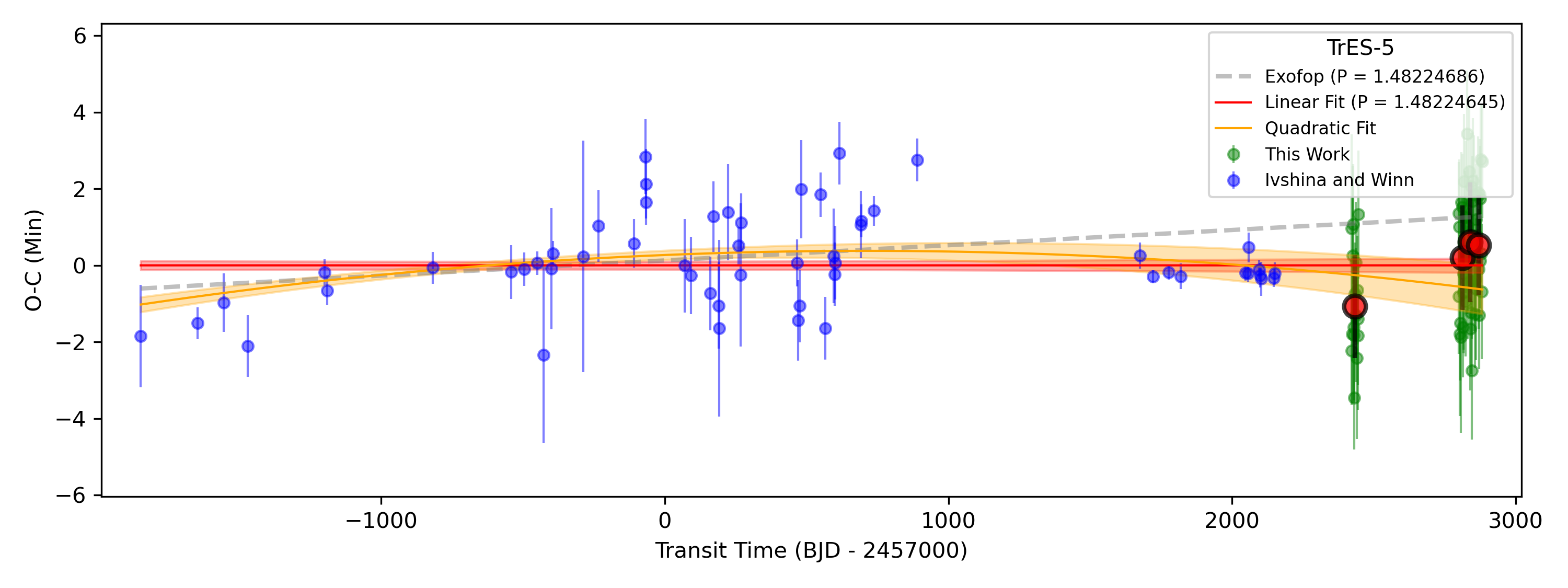}
\caption{Observed minus calculated transit time diagram for TrES-5. \label{fig:figfive}}
\end{figure*}

\textbf{TrES-5:} TrES-5 b is a hot Jupiter with a mass of $1.80 \pm 0.08 \: M_{J}$ and a radius of $1.20 \pm 0.02 \: R_{J}$ \citep{Maciejewski2016}. We fit 121 timing measurements, 66 of which were newly observed since the IW22 times were published. This gave us a $\dot{P}$ of $-13.91 \pm 2.77 \: ms \: yr^{-1}$, and a $Q_{*}^{'}$ of $(2.11 \pm 0.42) \times 10^{4}$. The quadratic fit was found to be better than the linear fit, with a $\Delta BIC$ of 19. TrES-5 was not specifically discussed by B20. However the stellar age estimate of $7.4 \pm 1.9 \: Gyr$ from \citet{Bonomo2017} gives a $Q_{IGW}^{'}$ of approximately $2.9 \times 10^{5}$ from figure 8 in B20, which is an order of magnitude above the value measured in this work. Table 4 in W24 gives an even higher range of $Q_{IGW}^{'}$ values for this system between $1.26-6.31 \times 10^{6}$.

\begin{figure*}[ht!]
\plotone{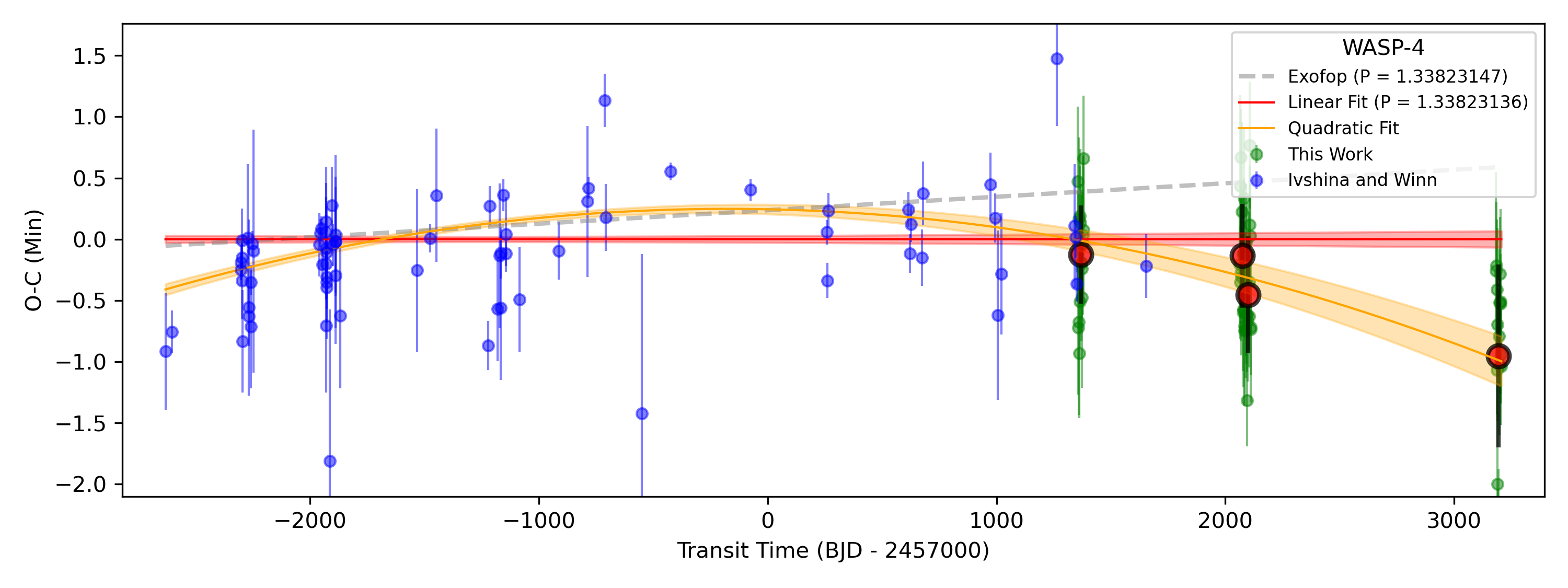}
\caption{Observed minus calculated transit time diagram for WASP-4. \label{fig:figsix}}
\end{figure*}

\textbf{WASP-4:} WASP-4 b is a hot Jupiter with a mass of $1.19 \pm 0.09 \: M_{J}$ and a radius of $1.32 \pm 0.04 \: R_{J}$ \citep{Bouma2019}. We fit 121 timing measurements, none of which were newly observed since the IW22 times were published. This gave us a $\dot{P}$ of $-6.39 \pm 0.53 \: ms \: yr^{-1}$, and a $Q_{*}^{'}$ of $(5.72 \pm 0.48) \times 10^{4}$. The quadratic fit was found to be significantly better than the linear fit, with a $\Delta BIC$ of 141. B20 found a $Q_{IGW}^{'}$ of $3.3-3.8 \times 10^{5}$, which is an order of magnitude above the value measured in this work. The stellar age estimate of $7.0 \pm 2.9 \: Gyr$ from \citet{Bonomo2017} gives an even higher range of $Q_{IGW}^{'}$ values between $5.0 \times 10^{5}-1.0 \times 10^{6}$ from table 4 in W24. However, through radial velocity measurement \citet{Bouma2020} found strong evidence that WASP-4 is accelerating towards Earth due to the wide orbit of a companion low-mass star, which is likely contributing to the apparent period decrease. We agree with this conclusion, and therefore disregard WASP-4 as a strong candidate for orbital decay.
 


\begin{figure*}[ht!]
\plotone{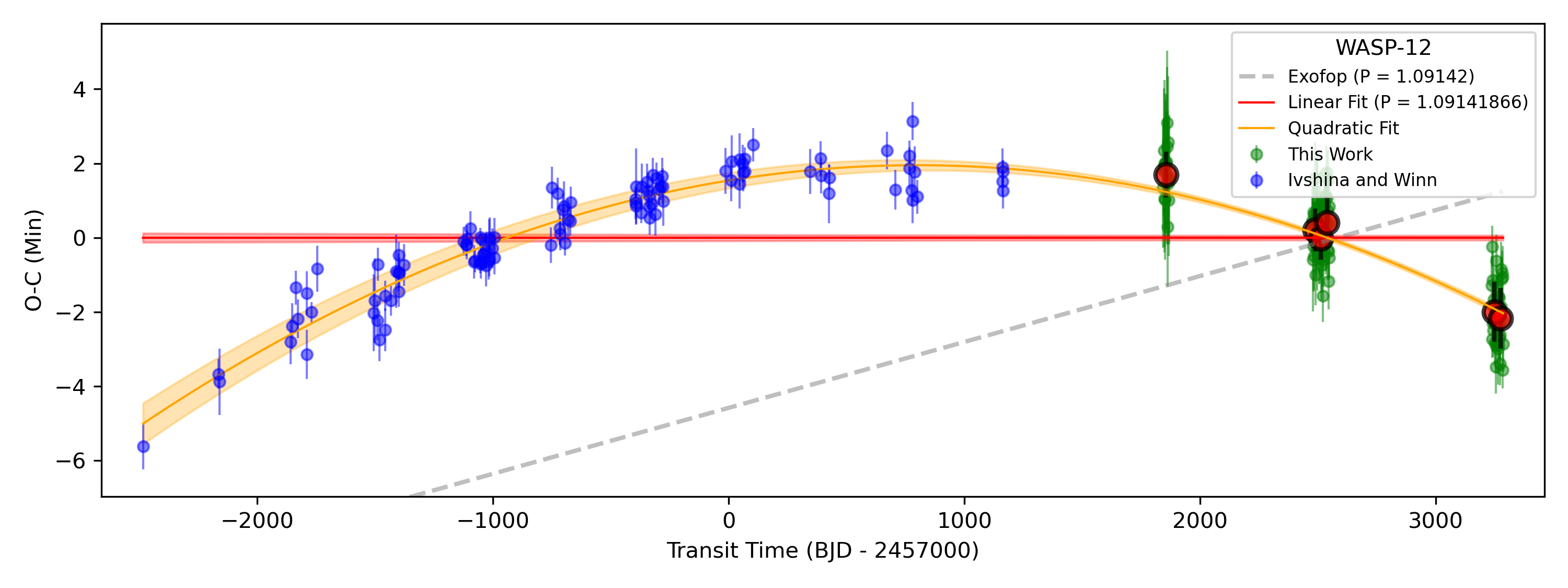}
\caption{Observed minus calculated transit time diagram for WASP-12. \label{fig:figseven}}
\end{figure*}

\textbf{WASP-12:} WASP-12 b is a hot Jupiter with a mass of $1.47 \pm 0.07 \: M_{J}$ and a radius of $1.90 \pm 0.06 \: R_{J}$ \citep{Collins2017}. We fit 189 timing measurements, 42 of which were newly observed since the IW22 times were published. This gave us a $\dot{P}$ of $-30.85 \pm 0.82 \: ms \: yr^{-1}$ and a $Q_{*}^{'}$ of $(1.64 \pm 0.06) \times 10^{5}$. The quadratic fit was found to be significantly better than the linear fit, with a $\Delta BIC$ of 1374. As this system has been studied in depth and we obtained results consistent with previous studies \citep[e.g.][]{Patra2017, Turner2021}, we do not discuss it further.

\begin{figure*}[ht!]
\plotone{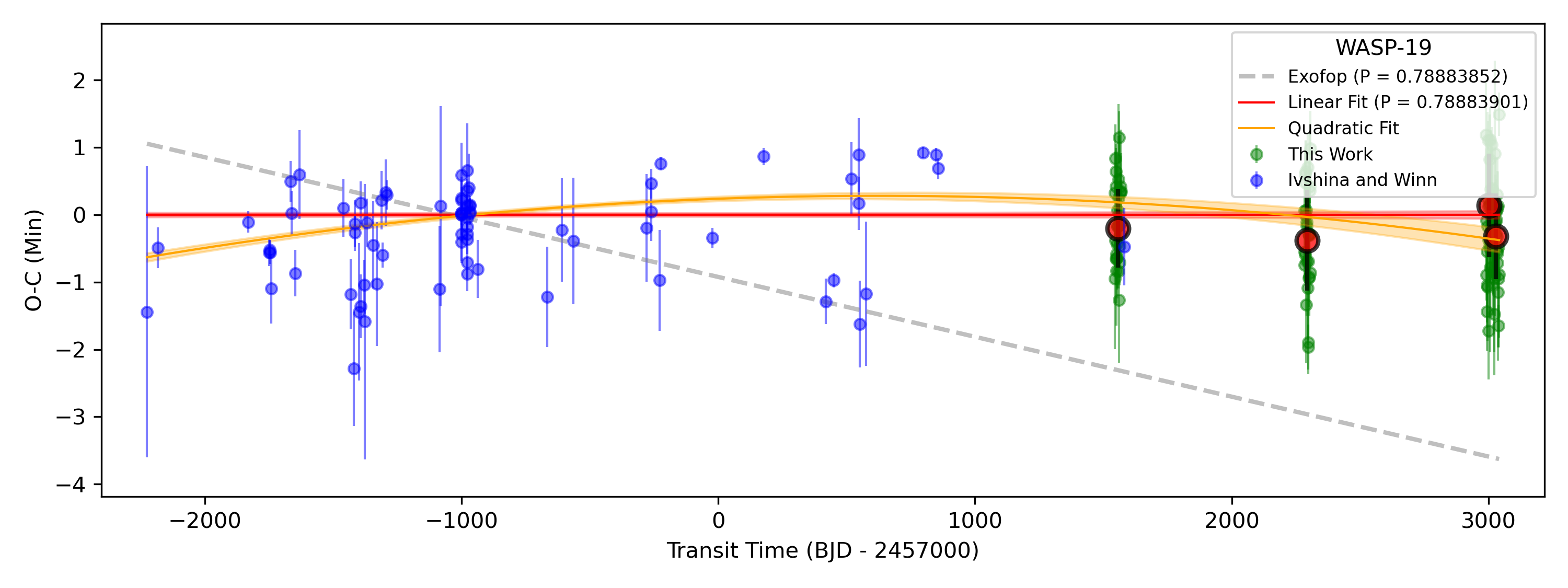}
\caption{Observed minus calculated transit time diagram for WASP-19. \label{fig:figeight}}
\end{figure*}

\textbf{WASP-19:} WASP-19 b is a hot Jupiter with a mass of $1.15 \pm 0.08 \: M_{J}$ and a radius of $1.42 \pm 0.05 \: R_{J}$ \citep{Cortes-Zuleta2020}. We fit 201 timing measurements, 62 of which were newly observed since the IW22 times were published. This gave us a $\dot{P}$ of $-3.90 \pm 0.37 \: ms \: yr^{-1}$, and a $Q_{*}^{'}$ of $(7.10 \pm 0.68) \times 10^{5}$. The quadratic fit was found to be significantly better than the linear fit, with a $\Delta BIC$ of 104. B20 found a $Q_{IGW}^{'}$ of $0.6-0.8 \times 10^{5}$, which is around an order of magnitude below the value measured in this work. B20 noted that they were uncertain whether wave-breaking occurs in the WASP-19 system due to varying estimates of the age of the system. Neglecting wave-breaking would have resulted in a higher $Q_{IGW}^{'}$. In fact, the stellar age estimate of $6.4_{-3.5}^{+4.1} \: Gyr$ from \citet{Cortes-Zuleta2020} gives a critical mass for wave breaking in the WASP-19 system above the mass of WASP-19 b from figure 9 from B20. Even more intriguing, table 4 in W24 gives a range of $Q_{IGW}^{'}$ between $2.0 - 7.9 \times 10^{5}$, which is consistent with the value measured in this work. However, similarly to CoRoT-2b, other sources have noted errors in the IW22 times of WASP-19b \citep{Adams2024}. When \citet{Adams2024} corrected these errors, they found insignificant evidence of orbital decay.

\begin{figure*}[ht!]
\plotone{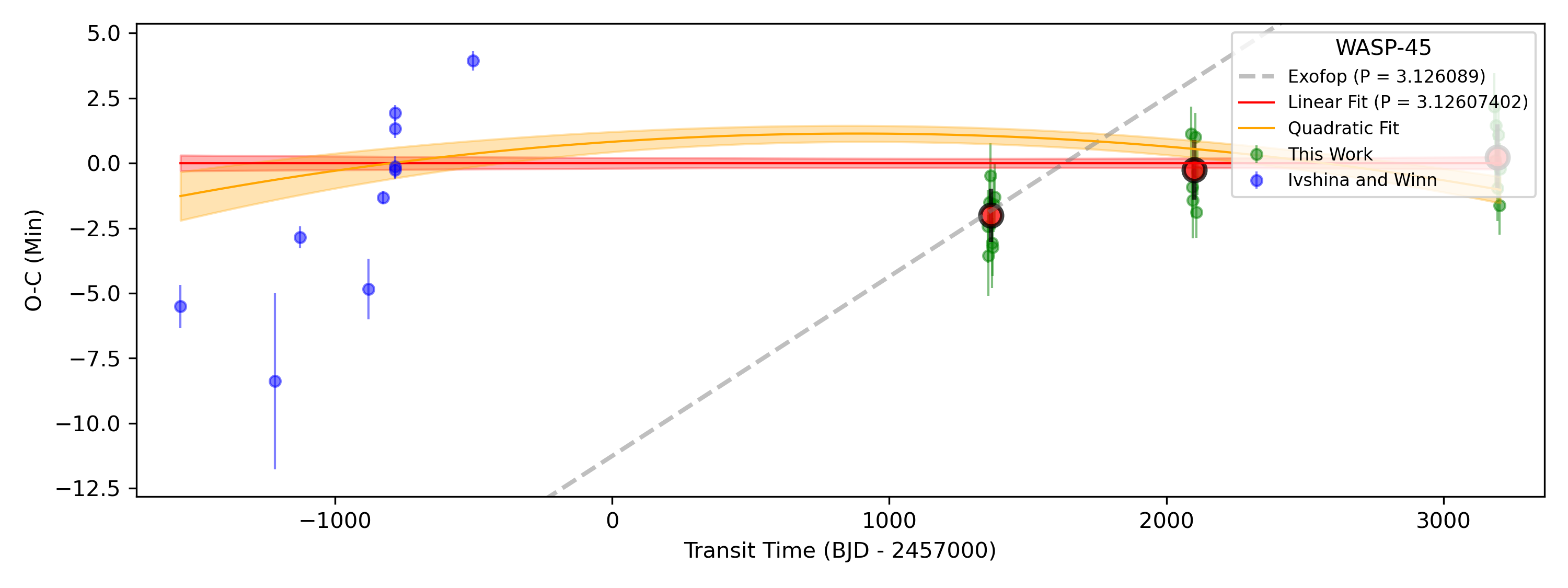}
\caption{Observed minus calculated transit time diagram for WASP-45. \label{fig:fignine}}
\end{figure*}

\textbf{WASP-45:} WASP-45 b is a hot Jupiter with a mass of $1.00 \pm 0.06 M_{J}$ and a radius of $0.99 \pm 0.04 \: R_{J}$ \citep{Ciceri2016}. We fit 24 timing measurements, none of which were newly observed since the IW22 times were published. This gave us a $\dot{P}$ of $-55.11 \pm 13.12 \: ms \: yr^{-1}$, and a $Q_{*}^{'}$ of $(3.31 \pm 0.79) \times 10^{2}$. The quadratic fit was found to be better than the linear fit, with a $\Delta BIC$ of 14. WASP-45 wasn't specifically discussed by B20, but the stellar age estimate of $12.7_{-5.3}^{+1.0} \: Gyr$ from \citet{Bonomo2017} gives a $Q_{IGW}^{'}$ of approximately $3.1 \times 10^{6}$ from figure 8 in B20, which is four orders of magnitude above the value measured in this work. Table 4 in W24 gives an even higher range of $Q_{IGW}^{'}$ values between $1.6-2.5 \times 10^{7}$. Despite the strong preference for a quadratic model over a linear model, it is unlikely that orbital decay is currently observable in the WASP-45 system as a result of the relatively large orbital period and high theoretical predictions of $Q_{IGW}^{'}$. Additionally, the data we had access to for WASP-45 is particularly sparse, the errors in transit times were likely under reported, and there is an additional star in the system whose existence will influence transit timing. Therefore, the $Q_{*}^{'}$ value measured in this work is likely not the result of any physical phenomenon.

\begin{figure*}[ht!]
\plotone{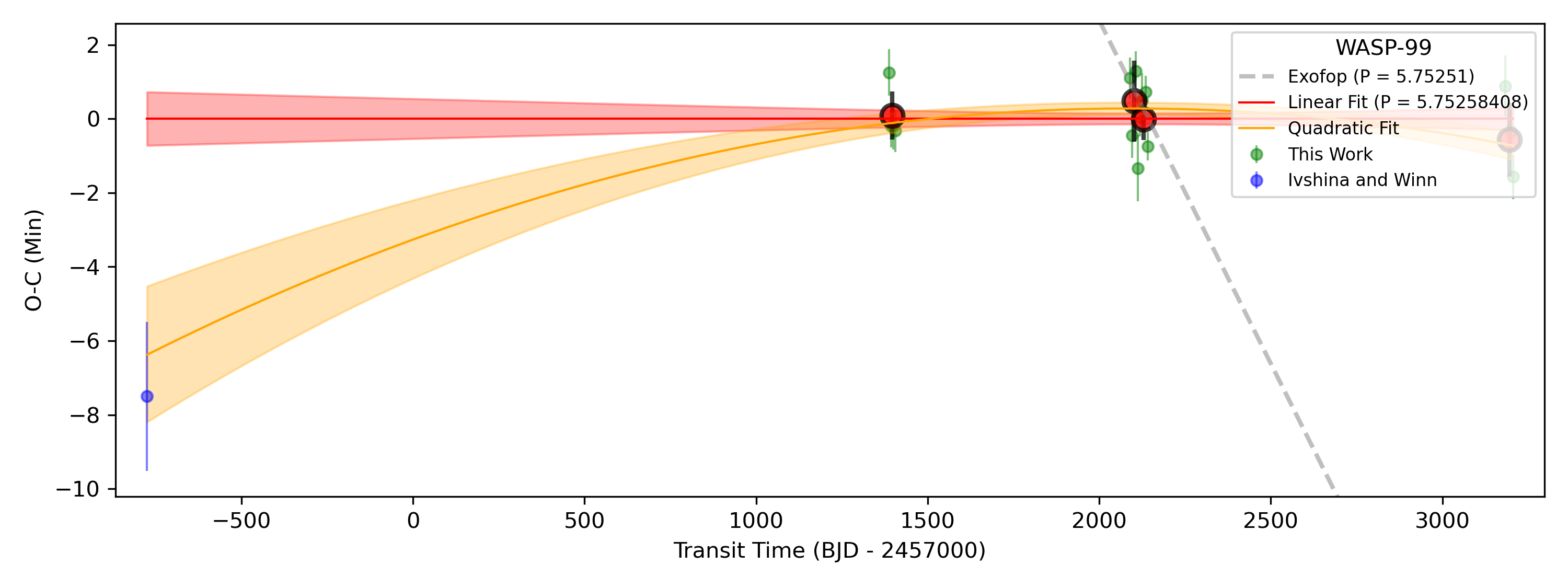}
\caption{Observed minus calculated transit time diagram for WASP-99. \label{fig:figten}}
\end{figure*}

\textbf{WASP-99:} WASP-99 b is a hot Jupiter with a mass of $2.78 \pm 0.13 M_{J}$ and a radius of $1.10 \pm 0.07 \: R_{J}$ \citep{Hellier2014}. We fit 16 timing measurements, 4 of which were newly observed since the IW22 times were published. This gave us a $\dot{P}$ of $-204.13 \pm 50.00 \: ms \: yr^{-1}$, and a $Q_{*}^{'}$ of $226.71 \pm 55.54$. The quadratic fit was found to be better than the linear fit, with a $\Delta BIC$ of 13. The goodness of fit of the quadratic model visibly relies on one transit time from IW22. This time represents the best fit $t_{0}$ of only two transits from ground-based transit surveys, which may influence the reliability of that transit time. However, evaluating the reliability of individual times is beyond the scope of this work. WASP-99 wasn't specifically discussed by B20. However, the stellar age estimate of $1.4_{-0.6}^{+1.1} \: Gyr$ from \citet{Hellier2014} gives a $Q_{IGW}$ of $2.0 \times 10^{7}$ from figure 8 in B20. The parameters of the WASP-99 system did not closely match any models from table 4 in W24 because of the uniquely high orbital period. In fact, it is unlikely that orbital decay is currently observable in the WASP-99 system as a result of this orbital period and the high theoretical prediction of $Q_{IGW}^{'}$. Furthermore, the $Q_{*}^{'}$ value measured in this work is likely the result of data sparsity and not any physical phenomenon because it is several orders of magnitude smaller than the expected $Q_{*}^{'}$ for any system.

\begin{figure*}[ht!]
\plotone{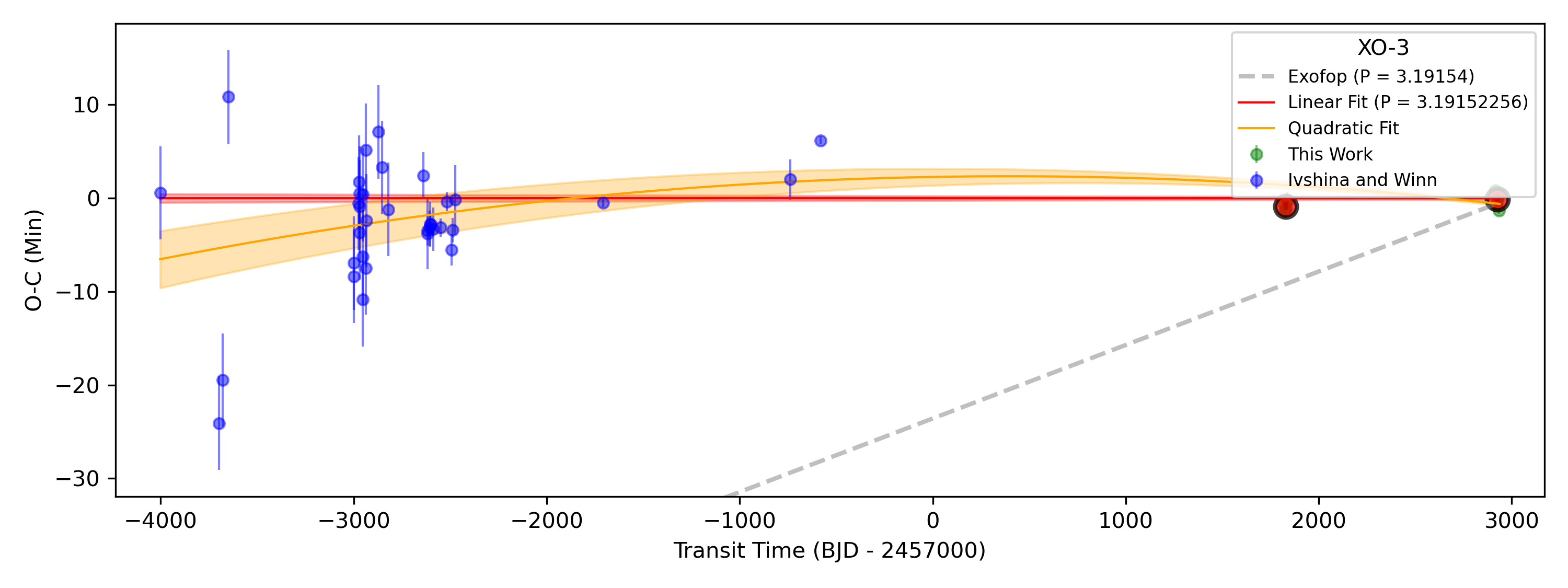}
\caption{Observed minus calculated transit time diagram for XO-3 b. \label{fig:figeleven}}
\end{figure*}

\textbf{XO-3:} XO-3 b is a hot Jupiter with a mass of $7.29 \pm 1.19 \: M_{J}$ and a radius of $1.41 \pm 0.12 R_{J}$ \citep{Stassun2017}. We fit 47 timing measurements, 6 of which were newly observed since the IW22 times were published. This gave us a $\dot{P}$ of $-64.18 \pm 7.06 \: ms \: yr^{-1}$, and a $Q_{*}^{'}$ of $(1.24 \pm 0.14) \times 10^{4}$. The quadratic fit was significantly better than the linear fit, with a $\Delta BIC$ of 79. XO-3 was not specifically discussed by B20, but the stellar age estimate of $2.6_{-1.1}^{+1.6} \: Gyr$ obtained by \citet{Worku2022} gives a $Q_{*}^{'}$ of approximately $2.2 \times 10^{6}$ from figure 8 in B20, which is two orders of magnitude above the value measured in this work. The parameters of the XO-3 system do not closely match any of the models from W24 because of the uniquely high planetary mass. There are three transit times that drive the quadratic fit away from the linear fit. The middle time represents a single transit observation by the Kuiper Telescope in Arizona \citep{Turner2017}. The first and third times represent single transits observed by the space-based Spitzer telescope \citep{Wong2014}. We note that the evidence for orbital decay is heavily dependent on the transit times from Spitzer. Specifically, if we remove the second such time, we find insignificant evidence for orbital decay, consistent with \citet{Wang2023}.

\clearpage

\begin{turnpage}
\begin{table*}
\centering
\caption{Model Parameters For \textbf{Notable Systems} \label{tab:tabtwo}}
\begin{tabular}{llllllll}
\hline
\multicolumn{8}{c}{Significant Systems} \\
\hline
\multicolumn{2}{c}{} & \multicolumn{2}{|c|}{Linear} & \multicolumn{3}{c|}{Quadratic} & \\
\hline
System & $N^{a}$ & $t_{0}$ & Period & $t_{0}$ & Period & $\frac{dP}{dE}$ & $\Delta BIC$ \\
&  & (BJD$^{b}$) & (days) & (BJD$^{b}$) & (days) & (days) \\
\hline
CoRoT-2 & $117$ & $-2762.468 \pm 1.58 \times 10^{-5}$ & $1.743 \pm 4.04 \times 10^{-8}$ & $-2762.468 \pm 1.75 \times 10^{-5}$ & $1.743 \pm 1.76 \times 10^{-7}$ & $(-2.78 \pm 0.12) \times 10^{-9}$ & $507$ \\
TrES-5 & $120$ & $-541.408 \pm 7.02 \times 10^{-5}$ & $1.482 \pm 4.71 \times 10^{-8}$ & $-541.408 \pm 6.99 \times 10^{-5}$ & $1.482 \pm 1.19 \times 10^{-7}$ & $(-6.53 \pm 1.30) \times 10^{-10}$ & $19$ \\
WASP-4 & $129$ & $-1195.484 \pm 1.49 \times 10^{-5}$ & $1.338 \pm 1.33 \times 10^{-8}$ & $-1195.484 \pm 1.70 \times 10^{-5}$ & $1.338 \pm 2.15 \times 10^{-8}$ & $(-2.71 \pm 0.23) \times 10^{-10}$ & $141$ \\
WASP-12 & $221$ & $2550.243 \pm 4.38 \times 10^{-5}$ & $1.091 \pm 1.61 \times 10^{-8}$ & $2550.243 \pm 4.34 \times 10^{-5}$ & $1.091 \pm 4.91 \times 10^{-8}$ & $(-1.07 \pm 0.03) \times 10^{-9}$ & $1374$ \\
WASP-19 & $189$ & $-597.287 \pm 1.68 \times 10^{-5}$ & $0.789 \pm 7.61 \times 10^{-9}$ & $-597.287 \pm 1.89 \times 10^{-5}$ & $0.789 \pm 1.61 \times 10^{-8}$ & $(-9.72 \pm 0.92) \times 10^{-11}$ & $104$ \\
WASP-45 & $31$ & $1339.144 \pm 1.19 \times 10^{-4}$ & $3.126 \pm 1.85 \times 10^{-7}$ & $1339.145 \pm 2.11 \times 10^{-4}$ & $3.126 \pm 2.70 \times 10^{-7}$ & $(-5.46 \pm 1.30) \times 10^{-9}$ & $14$ \\
WASP-99 & $16$ & $2135.796 \pm 1.01 \times 10^{-4}$ & $5.753 \pm 9.73 \times 10^{-7}$ & $2135.796 \pm 1.12 \times 10^{-4}$ & $5.753 \pm 9.82 \times 10^{-7}$ & $(-3.72 \pm 0.91) \times 10^{-8}$ & $13$ \\
XO-3 & $47$ & $2936.098 \pm 1.09 \times 10^{-4}$ & $3.192 \pm 1.37 \times 10^{-7}$ & $2936.097 \pm 1.18 \times 10^{-4}$ & $3.192 \pm 5.85 \times 10^{-7}$ & $(-6.49 \pm 0.71) \times 10^{-9}$ & $79$ \\
\hline
\multicolumn{8}{c}{\textbf{Systems With Any Evidence of a Period Decrease}} \\
\hline
\multicolumn{2}{c}{} & \multicolumn{2}{|c|}{Linear} & \multicolumn{3}{c|}{Quadratic} & \\
\hline
System & $N^{a}$ & $t_{0}$ & Period & $t_{0}$ & Period & $\frac{dP}{dE}$ & $\Delta BIC$ \\
&  & (BJD$^{b}$) & (days) & (BJD$^{b}$) & (days) & (days) \\
\hline
HAT-P-60 & $11$ & $1360.941 \pm 3.76 \times 10^{-4}$ & $4.795 \pm 1.51 \times 10^{-6}$ & $1360.940 \pm 5.58 \times 10^{-4}$ & $4.795 \pm 9.54 \times 10^{-6}$ & $(-9.70 \pm 5.15) \times 10^{-8}$ & $1$ \\
Kepler-1658 & $83$ & $-1994.074 \pm 2.73 \times 10^{-4}$ & $3.849 \pm 1.70 \times 10^{-6}$ & $-1994.075 \pm 3.99 \times 10^{-4}$ & $3.849 \pm 3.40 \times 10^{-6}$ & $(-1.92 \pm 0.62) \times 10^{-8}$ & $5$ \\
NGTS-6 & $56$ & $982.379 \pm 2.79 \times 10^{-4}$ & $0.882 \pm 3.64 \times 10^{-7}$ & $982.378 \pm 3.00 \times 10^{-4}$ & $0.882 \pm 1.55 \times 10^{-6}$ & $(-5.03 \pm 2.21) \times 10^{-9}$ & $1$ \\
TrES-3 & $276$ & $-2814.089 \pm 3.38 \times 10^{-5}$ & $1.306 \pm 1.18 \times 10^{-8}$ & $-2814.089 \pm 6.16 \times 10^{-5}$ & $1.306 \pm 8.02 \times 10^{-8}$ & $(-8.51 \pm 3.32) \times 10^{-11}$ & $1$ \\
WASP-5 & $76$ & $-2626.003 \pm 8.10 \times 10^{-5}$ & $1.628 \pm 4.26 \times 10^{-8}$ & $-2626.003 \pm 9.77 \times 10^{-5}$ & $1.628 \pm 1.69 \times 10^{-7}$ & $(-3.17 \pm 0.98) \times 10^{-10}$ & $5$ \\
WASP-114 & $12$ & $-332.263 \pm 2.05 \times 10^{-4}$ & $1.549 \pm 2.36 \times 10^{-7}$ & $-332.263 \pm 2.06 \times 10^{-4}$ & $1.549 \pm 1.40 \times 10^{-6}$ & $(-4.95 \pm 1.40) \times 10^{-9}$ & $9$ \\
\hline
\multicolumn{8}{l}{a. Number of transit times fit in this study.} \\
\multicolumn{8}{l}{b. Barycentric Julian Date minus 2457000 days.} \\
\multicolumn{8}{l}{(Entire table is available for download in machine readable format.)} \\
\end{tabular}
\end{table*}
\end{turnpage}  

\begin{table*}
\centering
\caption{Constraints on $Q_{*}^{'}$ For \textbf{Notable Systems} \label{tab:tabthree}}
\begin{tabular}{lllll}
\hline
\multicolumn{5}{c}{Significant Systems} \\
\hline
System & $\dot{P}$ & $\Delta BIC$ & $Q_{*}^{'}$ & $Q_{min}^{'}$ \\
& ($ms \: yr^{-1}$) &  &  & \\
\hline
CoRoT-2 & $-50.32 \pm 2.18$ & $507$ & $(6.33 \pm 0.27) \times 10^{3}$ & $5.60 \times 10^{3}$ \\
TrES-5 & $-13.91 \pm 2.77$ & $19$ & $(2.11 \pm 0.42) \times 10^{4}$ & $1.32 \times 10^{4}$ \\
WASP-4 & $-6.39 \pm 0.53$ & $141$ & $(5.72 \pm 0.48) \times 10^{4}$ & $4.59 \times 10^{4}$ \\
WASP-12 & $-30.85 \pm 0.82$ & $1374$ & $(1.64 \pm 0.04) \times 10^{5}$ & $1.51 \times 10^{5}$ \\
WASP-19 & $-3.89 \pm 0.37$ & $104$ & $(7.12 \pm 0.67) \times 10^{5}$ & $5.55 \times 10^{5}$ \\
WASP-45 & $-55.11 \pm 13.12$ & $14$ & $(3.31 \pm 0.79) \times 10^{2}$ & $1.93 \times 10^{2}$ \\
WASP-99 & $-204.13 \pm 50.00$ & $13$ & $(2.27 \pm 0.56) \times 10^{2}$ & $1.31 \times 10^{2}$ \\
XO-3 & $-64.18 \pm 7.06$ & $79$ & $(1.24 \pm 0.14) \times 10^{4}$ & $9.29 \times 10^{3}$ \\
\hline
\multicolumn{5}{c}{\textbf{Systems with any Evidence of a Period Decrease}} \\
\hline
System & $\dot{P}$ & $\Delta BIC$ & $Q_{*}^{'}$ & $Q_{min}^{'}$ \\
& ($ms \: yr^{-1}$) &  &  & \\
\hline
HAT-P-60 & $-638.60 \pm 339.14$ & $1$ & $(1.01 \pm 0.54) \times 10^{2}$ & $39.18$ \\
Kepler-1658 & $-157.75 \pm 50.65$ & $5$ & $(3.02 \pm 0.97) \times 10^{4}$ & $1.54 \times 10^{4}$ \\
NGTS-6 & $-179.92 \pm 79.22$ & $1$ & $(4.93 \pm 2.17) \times 10^{3}$ & $2.12 \times 10^{3}$ \\
TrES-3 & $-2.06 \pm 0.80$ & $1$ & $(1.75 \pm 0.68) \times 10^{5}$ & $8.11 \times 10^{4}$ \\
WASP-5 & $-6.14 \pm 1.91$ & $5$ & $(6.30 \pm 1.95) \times 10^{4}$ & $3.27 \times 10^{4}$ \\
WASP-114 & $-100.83 \pm 28.45$ & $9$ & $(1.19 \pm 0.34) \times 10^{4}$ & $6.51 \times 10^{3}$ \\
\hline
\end{tabular}
\end{table*}

\begin{table*}
\centering
\caption{Constraints on $Q_{*}^{'}$ For Systems Without Evidence of a Period Decrease \label{tab:tabfour}}
\begin{tabular}{lllll}
\hline
System & $\dot{P}$ & $\Delta BIC$ & $Q_{*}^{'}$ & $Q_{min}^{'}$ \\
& ($ms \: yr^{-1}$) &  &  & \\
\hline
CoRoT-1 & $6 \pm 6.92$ & $-3$ &   & $3.24 \times 10^{4}$ \\
CoRoT-5 & $-14 \pm 54.35$ & $-3$ & $(9 \pm 34.64) \times 10^{2}$ & $67.66$ \\
HAT-P-23 & $2 \pm 1.66$ & $-3$ &   & $4.63 \times 10^{5}$ \\
HAT-P-40 & $-69 \pm 66.15$ & $-1$ & $(1 \pm 1.25) \times 10^{3}$ & $3.40 \times 10^{2}$ \\
HAT-P-67 & $76 \pm 37.55$ & $0$ &   & $2.08 \times 10^{3}$ \\
HATS-18 & $33 \pm 3.95$ & $65$ &   & \\
K2-97 & $20466 \pm 14883.46$ & $-1$ &   & $93.23$ \\
K2-132 & $2771 \pm 23049.21$ & $-2$ &   & $2.28$ \\
K2-161 & $-98228 \pm 104621.42$ & $-1$ & $10 \pm 10.75$ & $2.42$ \\
KELT-1 & $2739 \pm 318.96$ & $64$ &   & \\
KELT-11 & $50 \pm 179.25$ & $-2$ &   & $2.22 \times 10^{2}$ \\
KELT-12 & $536 \pm 826.54$ & $-2$ &   & $37.26$ \\
KELT-16 & $8 \pm 4.53$ & $-1$ &   & $1.54 \times 10^{6}$ \\
Kepler-14 & $-275 \pm 206.64$ & $-2$ & $(1.1 \pm 0.82) \times 10^{3}$ & $3.35 \times 10^{2}$ \\
TOI-1181 & $285 \pm 29.99$ & $84$ &   & \\
TOI-1789 & $-27950 \pm 32481.32$ & $-3$ & $13 \pm 15.44$ & $3.10$ \\
TOI-2330 & $6544 \pm 3441.60$ & $0$ &   & $1.22 \times 10^{2}$ \\
TOI-2337 & $9828 \pm 1848.83$ & $24$ &   & \\
TOI-2787 & $-9412 \pm 48475.35$ & $-3$ & $(1 \pm 6.88) \times 10^{2}$ & $8.84$ \\
TOI-4379 & $-5004 \pm 28681.86$ & $-2$ & $(2 \pm 10.88) \times 10^{2}$ & $10.21$ \\
TOI-4436 & $768 \pm 399.13$ & $0$ &   & $1.73 \times 10^{4}$ \\
TOI-4603 & $2706 \pm 28907.44$ & $-2$ &   & $8.73$ \\
TOI-6029 & $332 \pm 36853.10$ & $-1$ &   & $5.95$ \\
TrES-1 & $4 \pm 2.18$ & $-1$ &   & $4.62 \times 10^{3}$ \\
WASP-1 & $26 \pm 5.40$ & $19$ &   & \\
WASP-2 & $2 \pm 3.80$ & $-3$ &   & $4.56 \times 10^{3}$ \\
WASP-3 & $2 \pm 2.34$ & $-4$ &   & $1.43 \times 10^{5}$ \\
WASP-6 & $4 \pm 3.94$ & $-2$ &   & $3.50 \times 10^{2}$ \\
WASP-18 & $0.9 \pm 0.49$ & $-1$ &   & $5.47 \times 10^{7}$ \\
WASP-33 & $7.9 \pm 0.51$ & $233$ &   & \\
WASP-43 & $2.2 \pm 0.40$ & $23$ &   & \\
WASP-71 & $40 \pm 26.44$ & $0$ &   & $2.88 \times 10^{4}$ \\
WASP-72 & $31 \pm 42.48$ & $-3$ &   & $1.97 \times 10^{5}$ \\
WASP-82 & $-66 \pm 123.44$ & $-2$ & $(1 \pm 2.36) \times 10^{4}$ & $1.90 \times 10^{3}$ \\
WASP-121 & $12 \pm 1.26$ & $87$ &   & \\
WASP-136 & $-274 \pm 171.90$ & $0$ & $(8 \pm 5.24) \times 10^{2}$ & $2.91 \times 10^{2}$ \\
WASP-161 & $-160 \pm 258.53$ & $-2$ & $(3 \pm 5.32) \times 10^{2}$ & $57.05$ \\
WASP-187 & $4 \pm 82.80$ & $-2$ &   & $1.37 \times 10^{3}$ \\
WASP-189 & $28 \pm 59.24$ & $-1$ &   & $3.38 \times 10^{3}$ \\
XO-1 & $-1 \pm 2.82$ & $-4$ & $(7 \pm 30.03) \times 10^{3}$ & $5.50 \times 10^{2}$ \\
\hline
\end{tabular}
\end{table*}

\clearpage

\section{Discussion} \label{sec:discussion}
 
\subsection{Implications of Orbital Decay Parameters} \label{subsec:implications}
\begin{figure*}[ht!]
\plotone{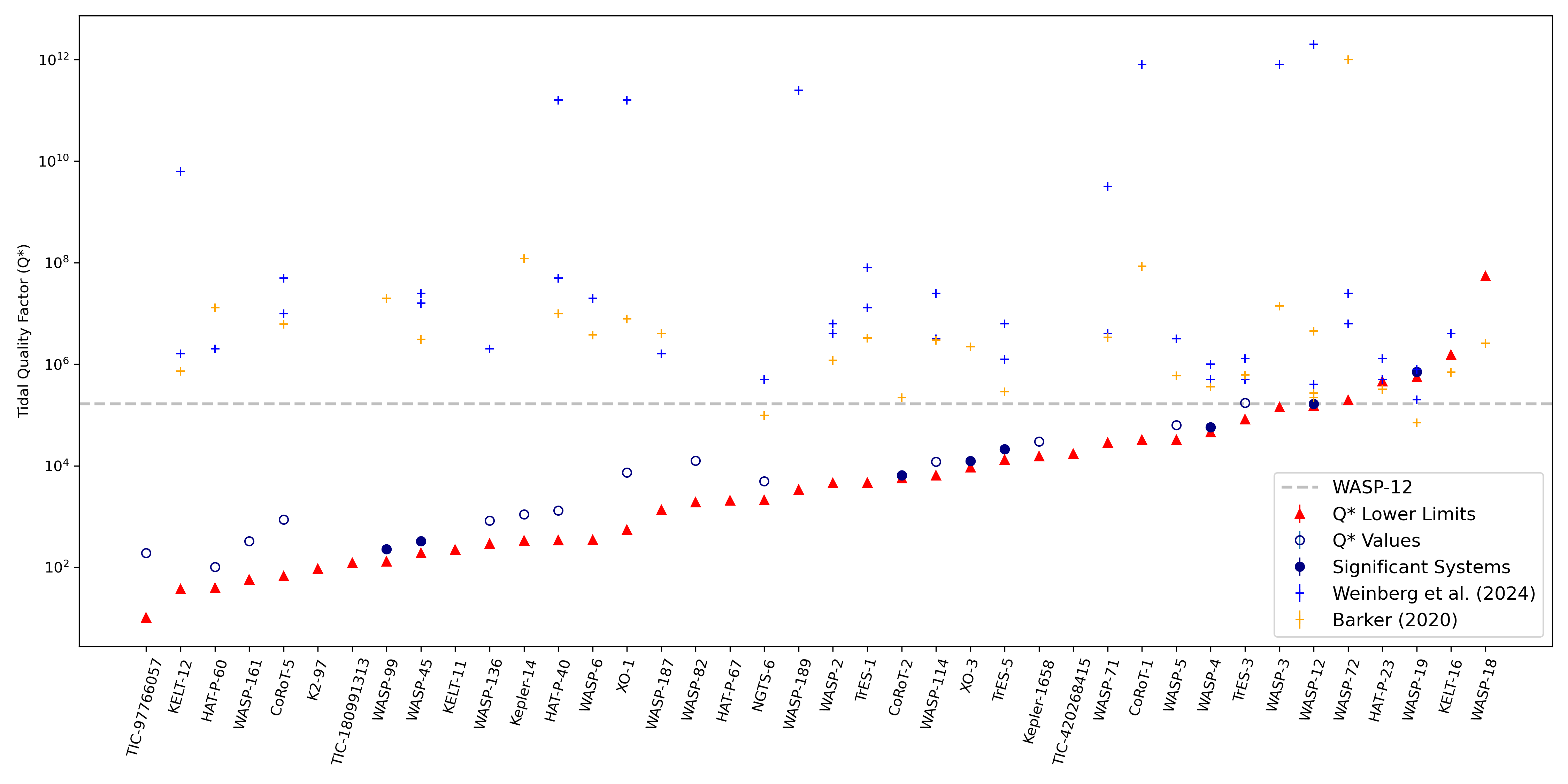}
\caption{Comparison between between our $Q_{*}^{'}$ values (circles), our $Q_{*}^{'}$ lower limits (red arrows), and theoretical estimates obtained by \citet{Barker2020} (orange crosses) and \citet{Weinberg2024} (blue crosses). Systems with two theoretical estimates from \citet{Weinberg2024} had high uncertainties in stellar ages or stellar ages in between two models, so we report lower and upper \citet{Weinberg2024} estimates for those systems. We highlight systems that show significant evidence of a deviation from a linear ephemeris as closed circles, while remaining systems are shown as open circles. Three \citet{Barker2020} values are displayed for WASP-12 (two are overlapping at around $2 \times 10^{5}$) because \citet{Barker2020} calculated $Q_{*}^{'}$ for three different stellar models of WASP-12 across different evolutionary states. \label{fig:figtwelve}}
\end{figure*}

Of our 8 significant systems, WASP-12 and WASP-19 had the best agreement with theory. Specifically, our $Q_{*}^{'}$ value for WASP-19 was consistent with estimates from W24 and strongly consistent with the non-wave breaking models of B20. It is predicted that main sequence systems with stellar masses between 0.5-1.1 $M_\odot$ and orbital periods of $\sim 1$ day should have $Q_{IGW}^{'}$ values of around $2 \times 10^{5}$, with $Q_{IGW}^{'} \propto P^{\frac{8}{3}}$, suggesting that 25-50\% of our 8 significant systems and 6 systems with any evidence of a period decrease should have $Q_{*}^{'}$ on the order of $10^{5}$ \citep{Barker2010, Essick2016, Barker2020}. The only systems in either of these categories with $Q_{*}^{'}$ on the order of $10^{5}$ were WASP-12 and WASP-19, with 5 systems having $Q_{*}^{'}$ values on the order of $10^{4}$. The inconsistencies between many of the other $Q_{*}^{'}$ values we obtained and theoretical predictions may be due to sparsity of data or errors in transit times from the literature. A comparison between our $Q_{*}^{'}$ values and lower limits, and theoretical values obtained by B20 and W24 is shown in figure \ref{fig:figtwelve}.

To provide conclusive evidence for orbital decay in any of the systems in the study, future studies would have to account for other mechanisms that result in apparent period decreases. A well-known mechanism is apsidal precession, which is a gradual shift of the orientation of an eccentric orbit around its star \citep{Patra2017}. Other mechanisms include the Rømer effect \citep{Bouma2019}, or an apparent decrease in period due to the acceleration of a planetary system towards Earth, and the Applegate effect \citep{Applegate1992}, which involves periodic changes in the structures of magnetically active stars that result in the transfer of angular momentum between the stars and their companions. Both apsidal precession and the Applegate effect would result in non-quadratic changes in the timing of transits, and the Rømer effect can be constrained because the wide orbiting companions that cause systems to accelerate towards Earth produce correlated signals in both transit timing and radial velocity data \citep{Wang2023}. Another common mechanism that causes non-quadratic changes in the timing of transits is the presence of hidden additional planets in the systems, which can be similarly modeled \citep{Wu2023}. It is generally understood that WASP-4 is undergoing the Rømer effect, as previously described \citep{Bouma2020}. 

In addition to mechanisms that change overall transit timing, there are various mechanisms that could interfere with the transit modeling process, resulting in inaccurate transit times. Large starspots, or any other stellar variability that has not been properly removed prior to fitting, can distort the shapes of transits, leading to spurious transit timing variations \citep{Silva-Valio2010}.

It is also important to note the known errors in the IW22 data. Since IW22 was a compilation from over 1000 sources, some inconsistencies in individual transit times have since been uncovered. These include duplicated transit times, individual transit times derived from multiple epochs of data that tend to have underestimated error bars, and incorrect conversions between various timing systems \citep{Adams2024}. Many recent studies have attempted to remove these errors. The identification of all of these errors is beyond the scope of this analysis. However, we have discussed the effects of these errors in the well-studied systems CoRoT-2, WASP-19, and XO-3. We consider the effects of these errors on our analysis in more detail in \S \ref{sec:results}. Additionally, for TrES-3, we omitted IW22 times with error bars that are known to be under-reported \citep{Maciejewski2024}. For WASP-161, we omitted one IW22 time with an unrealistically low error bar that seemed to be the sole reason for a nonlinear trend. For both systems, the removal of these times significantly decreased their $\Delta BIC$. In general, the longer the baseline of observation, the easier it is to confirm or reject orbital decay in a given system, despite errors in older transit times. Therefore, searches for orbital decay such as the one performed in this study will only become more complete and sensitive as more transit times are compiled. This bodes well for the future of such studies, especially since the methods of obtaining transit times are generally increasing in precision and an increasing number of errors are being caught.

We also note that there were 29 systems that showed an increase in orbital period, 8 of which had $\frac{dP}{dE}$ values were inconsistent with 0 by $3 \sigma$ and $\Delta BIC$ values greater than 10. These systems were TOI-2337, KELT-1, WASP-33, WASP-43, HATS-18, WASP-121, WASP-1, and TOI-1181. This could be due to astrophysical phenomena or measurement errors. Additionally, it could also be due to data sparsity, as in the case of KELT-1 which only had two effective epochs of data separated by 1000 days. A detailed analysis of these systems is beyond the scope of this study, though many of them would be intriguing targets for follow-up studies.

\subsection{Evolved Systems} \label{subsec:evolved}
We specifically flagged evolved systems ($T_{\mathrm{eff}} \lesssim 6000$ K, $log(g) \lesssim 4$) for further analysis, as evolved systems are suspected to have faster rates of orbital decay than main-sequence systems. The only evolved system that was flagged by our study as having significant evidence of orbital decay is WASP-99 ($T_{\mathrm{eff}} = 6079.00 \pm 131.58$, $log(g) = 4.03 \pm 0.08$). However, as previously discussed, the data for WASP-99 is particularly sparse and, as a result, the $Q_{*}^{'}$ value we measured is likely unphysical. Additionally, Kepler-1658 ($T_{\mathrm{eff}} = 5843.47 \pm 135.15$, $log(g) = 3.27 \pm 0.09$) was flagged as a system with any evidence of orbital decay by our study. \citet{Vissapragada2022} found significant evidence of orbital decay in the Kepler-1658 system, and we obtained a $Q_{*}^{'}$ consistent within $1 \sigma$ of the value obtained by \citet{Vissapragada2022} despite calculating a much lower $\Delta BIC$. This is likely a result of the fact that our analysis did not include transit times obtained by the Palomar Observatory, and thus the precision to which we could measure orbital decay was more limited in this study. In the future, new TESS data should be enough to confirm or refute orbital decay in this system.

In all other 11 evolved systems, we do not see any evidence of orbital decay. At first glance, it might seem that the stellar tidal quality factors of evolved systems would be lower than those of main-sequence systems \citep{Barker2020, Weinberg2024}. However, due to the effect of data sparsity and short baselines of observation, the constraints on the stellar tidal factors that we find here are not strong enough to distinguish between main-sequence and post-main-sequence systems. The median lower limit on $Q_{*}^{'}$ for this sample of evolved systems is 131 while the median lower limit for the full sample of systems studied here is 345. Though this qualitatively reflects that the tidal quality factor of the population of evolved systems is lower than that of hot Jupiter systems in general, both limits are orders of magnitudes lower than the $Q_{*}^{'}$ values expected in both samples.

In general, most of the evolved systems in this sample, and most of the systems in this study as a whole, don't have enough robust epochs of data to meaningfully constrain orbital decay. In order to significantly constrain orbital decay, collections of transit times spaced much further in time than the spacing between the transit times, or collections more precisely measured transit times, are needed. For example, the TOI-2337 system ($T_{\mathrm{eff}} = 4784.00 \pm 122.00$, $log(g) = 3.50 \pm 0.06$) was identified as likely experiencing rapid orbital decay as a result of its evolutionary state and its planet to star mass ratio \citep{Grunblatt2022}. However, the transit depth of this system is very small, so errors and variability in individual transit times dominated over longer term trends, such as orbital decay. Moreover, we only have two years of data for this system, which is insufficient to constrain $Q_{*}^{'}$ within an order of magnitude in which it has been predicted \citep{Barker2020, Weinberg2024}. Taken together, these factors make it unsurprising that orbital decay is poorly constrained in this system.

\subsection{Comparisons with Previous Studies} \label{subsec:comparisons}
\textbf{\citet{Ivshina2022}:} IW22 preformed a large transit timing survey for hundreds of hot Jupiter systems, but also flagged 10 systems that showed significant signs of long term period variations. Of those 10 systems, CoRoT-2, TrES-5, WASP-4, WASP-12, WASP-19, WASP-45, WASP-99, and XO-3 were flagged as significant targets by our study. We obtained $\dot{P}$ values within $1 \sigma$ of IW22's values for WASP-12 and WASP-19, and within $3 \sigma$ for TrES-5, WASP-4, and WASP-99. The remaining overlapping systems had at least 6 new transits observations by TESS since IW22 was published, with the exception of WASP-45, which had sparse data to begin with.

\textbf{\citet{Patra2020}:} \citet{Patra2020} preformed a search for orbital decay in 12 hot Jupiter systems. Of those systems, WASP 12 and WASP-19 were flagged as significant targets by our study. We obtained $\dot{P}$ values within $1 \sigma$ of Patra's values for WASP-12. We obtained $Q_{*}^{'}$ values within $3 \sigma$ of Patra's values for WASP-12 and within $5 \sigma$ for WASP-19.

\textbf{\citet{Wang2023}:} \citet{Wang2023} preformed a large search for orbital decay in 326 hot Jupiter systems, and flagged 18 systems that showed signs of decreasing periods. Of those 18 systems, CoRoT-2, TrES-5, WASP-4, WASP-12, WASP-19, WASP-45, and XO-3 were flagged as significant targets by our study. They employed both a $3 \sigma$ and $5 \sigma$ outlier rejection process for removing outlier transit times and calculated $\dot{P}$ values for each process. Here, we compare our values to Wang's $3 \sigma$ rejection values, as we used a similar $3 \sigma$ rejection process. We obtained $\dot{P}$ values within $1 \sigma$ of Wang's values for WASP-4, within $3 \sigma$ for TrES-5 and WASP-12, and within $5 \sigma$ for XO-3. Inconsistencies between our study and \citet{Wang2023} for the remaining systems highlights the sensitivities of our results to the specific details of the transit timing modeling process, especially when data is sparse, and remain important to explore.

\textbf{\citet{Adams2024}:} \citet{Adams2024} performed a comprehensive search for orbital decay in 43 hot Jupiter systems. Of the 4 systems for which they found $\Delta BIC > 30$, only WASP-12 was flagged as a significant target in our study. Of the remaining 39 systems in their study, CoRoT-2, WASP-5, and WASP-19 were flagged as significant targets in our study. We obtained $\dot{P}$ values within $1 \sigma$ of Adams' values for WASP-12, and within $3 \sigma$ for WASP-5. We obtained $Q_{*}^{'}$ values within $1 \sigma$ of Adams' values for WASP-5, within $3 \sigma$ for WASP-12. In their study, \citet{Adams2024} points out errors in literature transit times, specifically highlighting errors in the literature times of CoRoT-2 and WASP-19, as previously discussed above. This accounts for the inconsistencies between our values and Adams' values, and also highlights the caution needed when interpreting the trends in transit times for those two systems as well any systems for which transit times have been taken from heterogeneous surveys.

\section{Conclusions} \label{sec:conclusions}
The purpose of our study was to test models of tidal orbital decay for a selection of both main-sequence and evolved systems containing hot Jupiters, as such models predict that the efficiency of tidal dissipation is extremely sensitive to stellar age and other stellar characteristics. We achieved this by measuring transit times and fitting linear and quadratic ephemeris models to those transit times. We analyzed 51 systems containing confirmed hot Jupiters, and 3 systems that were flagged as TESS objects of interest. For every system, our linear models provided a new best fit constant period ephemeris, an essential part of searching for orbital decay as they are used to schedule new observations \citep{Jackson2023}. Out of the 54 systems in our study, 25 had negative $\dot{P}$ indicating a decreasing period. 8 of these systems had a $\dot{P}$ inconsistent with 0 by 3 standard deviations and a $\Delta BIC > 10$. We consider these 8 systems our most promising targets for orbital decay. Additionally, we consider 6 systems with $\dot{P}$ values inconsistent with 0 by at least 1 standard deviation and $\Delta BIC$ values between 1 and 10 to have potential evidence of orbital decay. The 8 promising targets are CoRoT-2, TrES-5, WASP-4, WASP-12, WASP-19, WASP-45, WASP-99, and XO-3, and the 6 systems with any evidence of orbital decay are HAT-P-60, Kepler-1658, NGTS-6, TrES-3, WASP-5, and WASP-114.

However, there are known errors with transit times that are available in the literature due to a variety of factors including underestimated error bars and incorrect timing conversions. Specifically, these errors are well documented for CoRoT-2, WASP-19, XO-3, and TrES-3, potentially making the detection of orbital decay in these systems less reliable. In fact, TrES-3 was originally one of our most promising targets, but the removal of only three transit times made its $\Delta BIC$ almost negligible. This highlights the importance of maintaining accurate transit timing databases as evaluation of individual transit times in population level analyses of exoplanet transits such as this study is challenging.

Similarly, we investigated a sample of evolved systems and found evidence for orbital decay in two systems, WASP-99 and Kepler-1658. However, our results for WASP-99 may be unreliable due to data sparsity. Our results for Kepler-1658 were consistent with previous studies, though our constraints are less precise because we do not use all available data.

We also calculated $Q_{*}^{'}$ values for every system with a negative $\dot{P}$, and lower limits on $Q_{*}^{'}$ for almost every system in the study to provide constraints on theoretical models, and compared them to various theoretical estimates of internal gravity wave-driven orbital decay. The median lower limit on $Q_{*}^{'}$ for evolved systems was marginally lower than it was for the sample of all systems, as predicted by theory, though both values are orders of magnitude lower than the theoretical expectations for these types of stars.

Understanding tidal interactions, and whether or not they lead to observable orbital decay, is key to understanding the evolution of all planetary systems and the internal structures of stars, which are challenging to probe. Tidal dissipation drives all tidal interactions, but theory is ahead of observation, so any constraints on $Q_{*}^{'}$ and orbital decay rate are welcome as models are continuously refined. The best way to do to provide these constraints is to consistently update linear ephemerides and provide new measurements of transit times, so data is readily available as the observational baselines of all planetary systems increase.

\begin{acknowledgements} 
We acknowledge the use of public TESS data from pipelines at the TESS Science Office and at the TESS Science Processing Operations Center. Resources supporting this work were provided by the NASA High-End Computing (HEC) Program through the NASA Advanced Supercomputing (NAS) Division at Ames Research Center for the production of the SPOC data products. S.G. acknowledges support by the National Aeronautics and Space Administration under grants 80NSSC23K0137 and 80NSSC23K0168. This research has made use of the Exoplanet Follow-up Observation Program website, which is operated by the California Institute of Technology under contract with the National Aeronautics and Space Administration under the Exoplanet Exploration Program. Funding for the TESS mission is provided by NASA's Science Mission Directorate. We would also like to acknowledge Mamaroneck High School's Original Science Research Program and the great work of Mr. Guido Garbarino and Mr. Alejandro Victoria in teaching research skills and fostering young scientists. This paper made use of the following software: \texttt{Astropy} \citep{Astropy2018}, \texttt{Astroquery} \citep{Ginsburg2019}, \texttt{Exoplanet} \citep{Foreman-Mackey2021}, and \texttt{PyMC} \citep{Salvatier2015}.
\end{acknowledgements} 

\section{Appendix}

\begin{figure*}[ht!]
\plotone{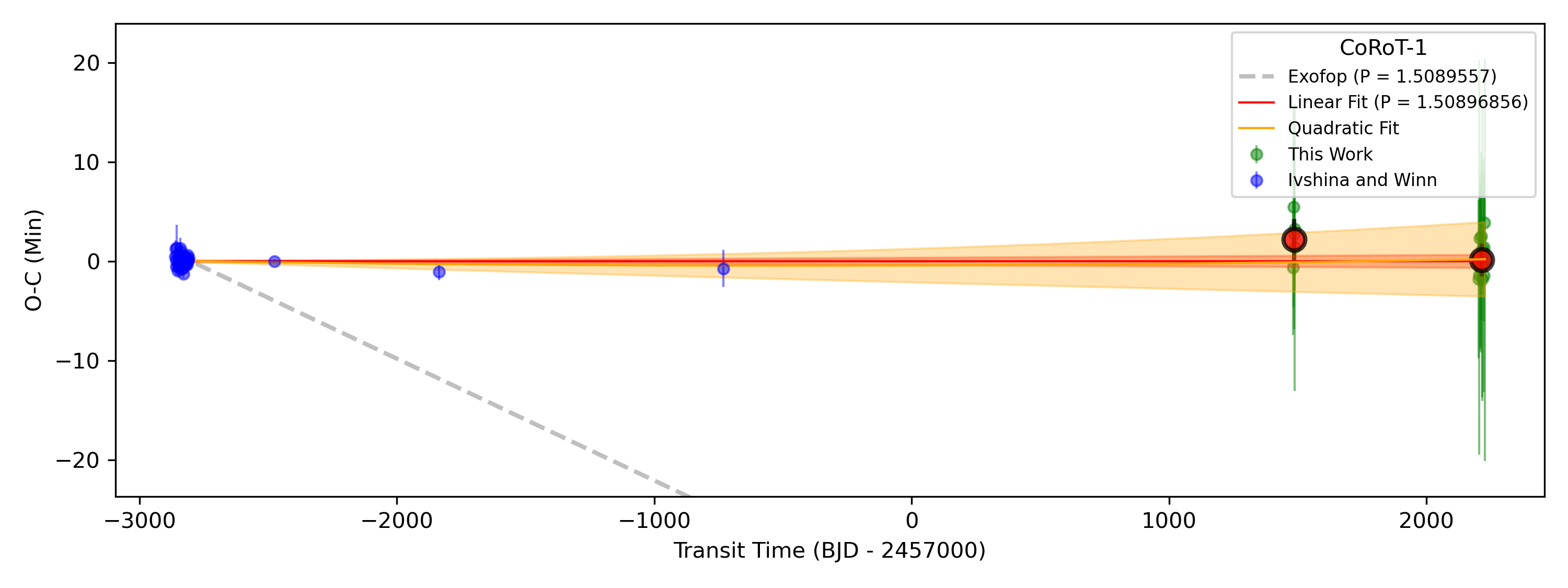}
\caption{Observed minus calculated transit time diagram for CoRoT-1, as in figure \ref{fig:figfour}. This figure is part of a figure set. The complete set (54 images) is available in the online journal \label{fig:figthirteen}}
\end{figure*}

To expand on our determination of orbital decay parameters, we include figure \ref{fig:figthirteen} which shows the observed minus calculated (O-C) transit time diagram for CoRoT-1 as an example of all of the systems studied here. Figure \ref{fig:figthirteen} is part of a figure set, which contains O-C transit time diagrams for every system in this study.

Additionally, we include Table \ref{tab:tabfive} which contains the first 20 specific transit times we measured. The times are reported in BJD - 2457000, which is using the standard timing format of TESS data, and the epoch numbers are calculated using equation \ref{eq:eqfour} with $t_{0}$ representing the first TESS transit time we measured for each system. The complete table, which contains the specific transit times of every system in this study, is available in its entirety in machine readable format.

\figsetstart
\figsetnum{13}
\figsettitle{Observed Minus Calculated Transit Time Diagrams}

\figsetgrpstart
\figsetgrpnum{13.1}
\figsetgrptitle{O-C Diagram of CoRoT-1}
\figsetplot{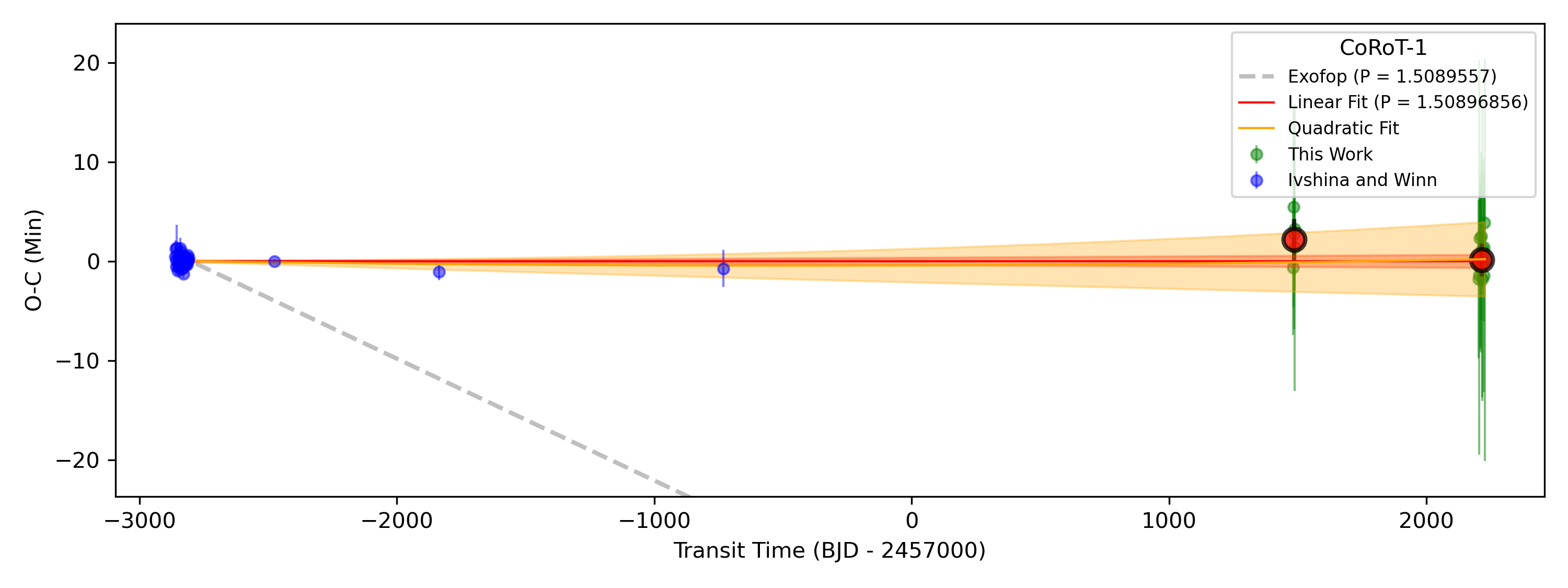}
\figsetgrpnote{Observed minus calculated transit time diagram for system listed in legend.}
\figsetgrpend

\figsetgrpstart
\figsetgrpnum{13.2}
\figsetgrptitle{O-C Diagram of CoRoT-2}
\figsetplot{CoRoT-2.png}
\figsetgrpnote{Observed minus calculated transit time diagram for system listed in legend.}
\figsetgrpend

\figsetgrpstart
\figsetgrpnum{13.30}
\figsetgrptitle{O-C Diagram of TrES-5}
\figsetplot{TrES-5.png}
\figsetgrpnote{Observed minus calculated transit time diagram for system listed in legend.}
\figsetgrpend

\figsetgrpstart
\figsetgrpnum{13.31}
\figsetgrptitle{O-C Diagram of WASP-1}
\figsetplot{WASP-1.png}
\figsetgrpnote{Observed minus calculated transit time diagram for system listed in legend.}
\figsetgrpend

\figsetgrpstart
\figsetgrpnum{13.32}
\figsetgrptitle{O-C Diagram of WASP-2}
\figsetplot{WASP-2.png}
\figsetgrpnote{Observed minus calculated transit time diagram for system listed in legend.}
\figsetgrpend

\figsetgrpstart
\figsetgrpnum{13.33}
\figsetgrptitle{O-C Diagram of WASP-3}
\figsetplot{WASP-3.png}
\figsetgrpnote{Observed minus calculated transit time diagram for system listed in legend.}
\figsetgrpend

\figsetgrpstart
\figsetgrpnum{13.34}
\figsetgrptitle{O-C Diagram of WASP-4}
\figsetplot{WASP-4.png}
\figsetgrpnote{Observed minus calculated transit time diagram for system listed in legend.}
\figsetgrpend

\figsetgrpstart
\figsetgrpnum{13.35}
\figsetgrptitle{O-C Diagram of WASP-5}
\figsetplot{WASP-5.png}
\figsetgrpnote{Observed minus calculated transit time diagram for system listed in legend.}
\figsetgrpend

\figsetgrpstart
\figsetgrpnum{13.36}
\figsetgrptitle{O-C Diagram of WASP-6}
\figsetplot{WASP-6.png}
\figsetgrpnote{Observed minus calculated transit time diagram for system listed in legend.}
\figsetgrpend

\figsetgrpstart
\figsetgrpnum{13.37}
\figsetgrptitle{O-C Diagram of WASP-12}
\figsetplot{WASP-12.png}
\figsetgrpnote{Observed minus calculated transit time diagram for system listed in legend.}
\figsetgrpend

\figsetgrpstart
\figsetgrpnum{13.38}
\figsetgrptitle{O-C Diagram of WASP-18}
\figsetplot{WASP-18.png}
\figsetgrpnote{Observed minus calculated transit time diagram for system listed in legend.}
\figsetgrpend

\figsetgrpstart
\figsetgrpnum{13.39}
\figsetgrptitle{O-C Diagram of WASP-19}
\figsetplot{WASP-19.png}
\figsetgrpnote{Observed minus calculated transit time diagram for system listed in legend.}
\figsetgrpend

\figsetgrpstart
\figsetgrpnum{13.40}
\figsetgrptitle{O-C Diagram of WASP-33}
\figsetplot{WASP-33.png}
\figsetgrpnote{Observed minus calculated transit time diagram for system listed in legend.}
\figsetgrpend

\figsetgrpstart
\figsetgrpnum{13.41}
\figsetgrptitle{O-C Diagram of WASP-43}
\figsetplot{WASP-43.png}
\figsetgrpnote{Observed minus calculated transit time diagram for system listed in legend.}
\figsetgrpend

\figsetgrpstart
\figsetgrpnum{13.42}
\figsetgrptitle{O-C Diagram of WASP-45}
\figsetplot{WASP-45.png}
\figsetgrpnote{Observed minus calculated transit time diagram for system listed in legend.}
\figsetgrpend

\figsetgrpstart
\figsetgrpnum{13.43}
\figsetgrptitle{O-C Diagram of WASP-71}
\figsetplot{WASP-71.png}
\figsetgrpnote{Observed minus calculated transit time diagram for system listed in legend.}
\figsetgrpend

\figsetgrpstart
\figsetgrpnum{13.44}
\figsetgrptitle{O-C Diagram of WASP-72}
\figsetplot{WASP-72.png}
\figsetgrpnote{Observed minus calculated transit time diagram for system listed in legend.}
\figsetgrpend

\figsetgrpstart
\figsetgrpnum{13.45}
\figsetgrptitle{O-C Diagram of WASP-82}
\figsetplot{WASP-82.png}
\figsetgrpnote{Observed minus calculated transit time diagram for system listed in legend.}
\figsetgrpend

\figsetgrpstart
\figsetgrpnum{13.46}
\figsetgrptitle{O-C Diagram of WASP-99}
\figsetplot{WASP-99.png}
\figsetgrpnote{Observed minus calculated transit time diagram for system listed in legend.}
\figsetgrpend

\figsetgrpstart
\figsetgrpnum{13.47}
\figsetgrptitle{O-C Diagram of WASP-114}
\figsetplot{WASP-114.png}
\figsetgrpnote{Observed minus calculated transit time diagram for system listed in legend.}
\figsetgrpend

\figsetgrpstart
\figsetgrpnum{13.48}
\figsetgrptitle{O-C Diagram of WASP-121}
\figsetplot{WASP-121.png}
\figsetgrpnote{Observed minus calculated transit time diagram for system listed in legend.}
\figsetgrpend

\figsetgrpstart
\figsetgrpnum{13.49}
\figsetgrptitle{O-C Diagram of WASP-136}
\figsetplot{WASP-136.png}
\figsetgrpnote{Observed minus calculated transit time diagram for system listed in legend.}
\figsetgrpend

\figsetgrpstart
\figsetgrpnum{13.50}
\figsetgrptitle{O-C Diagram of WASP-161}
\figsetplot{WASP-161.png}
\figsetgrpnote{Observed minus calculated transit time diagram for system listed in legend.}
\figsetgrpend

\figsetgrpstart
\figsetgrpnum{13.51}
\figsetgrptitle{O-C Diagram of WASP-187}
\figsetplot{WASP-187.png}
\figsetgrpnote{Observed minus calculated transit time diagram for system listed in legend.}
\figsetgrpend

\figsetgrpstart
\figsetgrpnum{13.52}
\figsetgrptitle{O-C Diagram of WASP-189}
\figsetplot{WASP-189.png}
\figsetgrpnote{Observed minus calculated transit time diagram for system listed in legend.}
\figsetgrpend

\figsetgrpstart
\figsetgrpnum{13.53}
\figsetgrptitle{O-C Diagram of XO-1}
\figsetplot{XO-1.png}
\figsetgrpnote{Observed minus calculated transit time diagram for system listed in legend.}
\figsetgrpend

\figsetgrpstart
\figsetgrpnum{13.54}
\figsetgrptitle{O-C Diagram of XO-3}
\figsetplot{XO-3.png}
\figsetgrpnote{Observed minus calculated transit time diagram for system listed in legend.}
\figsetgrpend

\figsetend

\begin{table*}
\centering
\caption{Transit Timing Measurements \label{tab:tabfive}}
\begin{tabular}{lll}
\hline
System & Epoch$^{a}$ & Transit Time \\
 &  & BJD$^{b}$ \\
\hline
TrES-3 & 0 & $1984.83927 \pm 0.00112$ \\
TrES-3 & 1 & $1986.14561 \pm 0.00030$ \\
TrES-3 & 2 & $1987.45133 \pm 0.00060$ \\
TrES-3 & 3 & $1988.75773 \pm 0.00072$ \\
TrES-3 & 4 & $1990.06467 \pm 0.00067$ \\
TrES-3 & 5 & $1991.36988 \pm 0.00064$ \\
TrES-3 & 6 & $1992.67723 \pm 0.00055$ \\
TrES-3 & 7 & $1993.98268 \pm 0.00128$ \\
TrES-3 & 8 & $1995.28926 \pm 0.00042$ \\
TrES-3 & 10 & $1997.90161 \pm 0.00031$ \\
TrES-3 & 11 & $1999.20704 \pm 0.00034$ \\
TrES-3 & 12 & $2000.51439 \pm 0.00044$ \\
TrES-3 & 13 & $2001.81869 \pm 0.00067$ \\
TrES-3 & 14 & $2003.12552 \pm 0.00055$ \\
TrES-3 & 15 & $2004.43225 \pm 0.00025$ \\
TrES-3 & 16 & $2005.73697 \pm 0.00061$ \\
TrES-3 & 17 & $2007.04444 \pm 0.00041$ \\
TrES-3 & 18 & $2008.35061 \pm 0.00067$ \\
TrES-3 & 20 & $2010.96387 \pm 0.00080$ \\
TrES-3 & 21 & $2012.26977 \pm 0.00049$ \\
\hline
\multicolumn{3}{l}{a. Calculated using equation \ref{eq:eqfour} with $t_{0}$ representing the first TESS transit time we measured for each system.} \\
\multicolumn{3}{l}{b. Barycentric Julian Date minus 2457000 days.} \\
\multicolumn{3}{l}{(Entire table is available for download in machine readable format.)} \\
\end{tabular}
\end{table*}

\bibliography{PaperBib}{}
\bibliographystyle{aasjournal}

\end{document}